\documentclass[postprint,5p,times]{elsarticle}

\usepackage{nomencl}
\usepackage{multicol}
\usepackage{multirow}
\usepackage{xcolor}
\usepackage[utf8]{inputenc}
\usepackage{enumitem}
\usepackage{nicefrac}
\usepackage{todonotes}
\usepackage{hyperref}

\usepackage{tikz}
\newcommand\copyrighttext{%
  \footnotesize $^1$ This is the accepted version of the manuscript. It does not include publisher value-added contributions. \copyright{}2020. This manuscript version is made available under the CC-BY-NC-ND 4.0 license http://creativecommons.org. The original journal article is available at: \href{https://doi.org/10.1016/j.apenergy.2020.114523}{https://doi.org/10.1016/j.apenergy.2020.114523}}
\newcommand\copyrightnotice{%
\begin{tikzpicture}[remember picture,overlay]
\node[anchor=south,yshift=30pt] at (current page.south) {\parbox{\dimexpr\textwidth-\fboxsep-\fboxrule\relax}{\copyrighttext}};;% {\fbox{\parbox{\dimexpr\textwidth-\fboxsep-\fboxrule\relax}{\copyrighttext}}};
\end{tikzpicture}%
}

\makenomenclature

\usepackage{amsmath}
\usepackage{graphicx}
\usepackage{amssymb}
\usepackage{lineno}

\journal{Applied Energy$^1$}

\begin{document}

\begin{frontmatter}


\title{How regional differences in cost of capital influence the optimal design of power systems}%Implications of regional differences in cost of capital on the optimal design of power systems with ambitious CO$_2$ reduction targets}
%\title{why local differences in capital costs need to be considered in power system modeling}



\author[dlr]{Bruno~U.~Schyska\corref{bruno}}
\author[fias]{Alexander~Kies}
\address[dlr]{DLR-Institute of Networked Energy Systems, Carl-von-Ossietzky-Str. 15, 26129 Oldenburg, Germany}
\address[fias]{Frankfurt Institute for Advanced Studies, Goethe University Frankfurt, Ruth-Moufang-Str. 1, 60438 Frankfurt am Main, Germany}
\cortext[bruno]{corresponding author: Bruno Schyska (bruno.schyska@dlr.de)}

\begin{abstract}
In order to reduce greenhouse gas emissions of the power sector, high shares of renewable power sources need to be integrated into existing systems. This will require vast amounts of investments. Cost of the capital needed for these investments are unevenly distributed among European regions. They show a clear North-South and West-East divide, which has not exhibited significant signs of narrowing in recent years. Power system studies investigating a continent-wide European power system, however, usually assume homogeneous cost of capital.

The objective of this paper is to investigate how regional differences in cost of capital affect the result of these studies with respect to the optimal power system design. Our analysis is based on power system optimization with inhomogeneous cost of capital in Europe. We find that assuming homogeneous cost of capital leads to estimates on the levelized costs of electricity in a highly renewable European power system, which are too conservative. The optimal system design is significantly affected compared to an inhomogeneous scenario. In particular, we show that inhomogeneous cost of capital favors overall wind power deployment in the case of Europe, while the investment in solar power decreases.
\end{abstract}

\begin{keyword}
power system analysis \sep locational marginal price \sep WACC \sep investment \sep renewables \sep climate change mitigation


\end{keyword}

\end{frontmatter}

\copyrightnotice

%\linenumbers

\section{Introduction}
%-----------------------------------------------------

In order to fulfill the goals of the Paris climate agreement \cite{paris2015}, Europe must meet ambitious CO$_2$ reduction targets. In this context, renewable power sources play a major role. Due to their growing cost competitiveness and vast potential they offer an efficient way to decarbonize the electricity sector. Their relevance even increases, because other sectors besides electricity such as heat and transportation need to be decarbonized as well \cite{brown2018synergies}, which is likely to be done via electrification \cite{usher2012critical}. The transition towards a power system based on renewables is subject of intense research and various solutions have been proposed to ease the large-scale integration of intermittent renewable generation sources: One important aspect is the mixing of different renewable resources to profit from the smoothing effect. This smoothing effect has been investigated for various combinations, e.g. solar photovoltaics (PV) and hydro reservoir storage \cite{ming2017optimizing}, wind and solar power \cite{heide2010seasonal, thomaidis2016optimal}, solar and hydro power \cite{francois2016complementarity}, PV and run-off-river \cite{jurasz2017integrating}, solar, wind and pumped hydro storage \cite{jurasz2017modeling, jurasz2018large}, wind power and concentrated solar power \cite{santos2015combining}, and even small scale hybrid energy systems \cite{jurasz2018impact}. The smoothing effect can best be utilized in large scale interconnected systems. However, these systems require the enhancement of the transmission grid \cite{rodriguez2014transmission}. \citet{kies2016curtailment} for instance found that insufficient transmission capacities might cause large amounts of renewable generation to be curtailed. Consequently, \citet{cao2018incorporating} suggested to incorporate transmission bottlenecks directly into energy system models. Similar to the optimal mix of generation sources, there is a well-known interaction between transmission and storage \cite{steinke2013grid}. \citet{weitemeyer2015integration} found that storage devices -- especially small units with high efficiencies such as lithium-ion batteries \cite{dunn2011electrical} -- are beneficial for a renewable share of more than 50\% in Germany. For higher shares seasonal storage devices and the expansion of the transmission grid become more important \cite{weitemeyer2016european}. To a small extend over-capacities can reduce the need for both storage and transmission \cite{heide2011reduced}. Furthermore, options to modify the demand side have been investigated \cite{palensky2011demand, zerrahn2015representation}. \citet{kies2016demand} found that demand-side management can balance generation-side fluctuations for a renewable share of up to 65\% in Europe. More recently, the establishment of so-called system-friendly renewables such as wind turbines \cite{hirth2016system} or PV modules \cite{chattopadhyay2017impact} that are designed to resemble load patterns or the use of vehicle-to-grid technologies \cite{lund2008integration} have been proposed.

In the European Union, renewable shares are growing rapidly. This requires vast amounts of capital for investments. Contrary to conventional generation, where generation and operation cost (fuel costs, maintenance, etc.) is crucial, renewable energy sources like wind and solar PV require large upfront investments, while fuel costs are non-existent. In addition, most mentioned integration options are capital-intensive or might even require building up entirely new infrastructures such as demand-side management with smart meters or electric vehicles and their charging infrastructure. This increases the need for investments even further. The European Commission estimates necessary investments of 379 billion Euro p.a. in the European Union after 2021 \cite{europeancommission}. Zappa et al. \cite{zappa2019100} estimate the cost for building a 100\% renewable European power system until 2050 to be 560 billion Euro p.a.. These costs are to a large extent driven by the cost of the capital needed to make investments. According to \citet{noothout2016diacore}, conditions to invest in renewables vary considerably between regions in Europe. These variations are reflected in different rates at which investors can raise funds. \citet{noothout2016diacore} measure these rates as the \emph{weighted average cost of capital} (WACC). Regional differences in WACC mainly originate from different financing and tariff-related risks for renewables \cite{noothout2016diacore}. \citet{egli2018dynamic} reported that financing conditions for solar PV and wind power projects in Germany have improved significantly over the last years contributing to the reduction of the levelized cost of electricity from these two resources.

In this context, \citet{klessmann2013policy} suggested to reduce the capital needs for renewables by applying political means to reduce financing risks in the renewables sector. \citet{temperton2016reducing} proposed to establish a transnational European facility in order to reduce the cost for financing renewable projects in Europe. And \citet{kitzing2012renewable} showed that a convergence of policies to support renewables could be observed, which might spur a slight convergence of WACC in the future. However, \citet{brueckmann} found that no tightening of this WACC gap has occurred in recent years. An interaction between the WACC and the price for CO$_2$ emissions has been found by \citet{hirth2016role}. Accordingly, higher WACC require higher CO$_2$ prices to achieve the same reduction target.

Existing power system studies -- such as the studies mentioned above -- often apply power system optimization to find the cost-optimal system while keeping greenhouse gas emissions below a certain threshold. From this optimization, the following information can be obtained: What does the optimal system design look like and where should generation, storage and transmission capacity be erected? What is the levelized cost of electricity? These studies, however, often do not consider regional differences in the cost of capital: \citet{schlachtberger2017benefits}, for instance, assumed spatially homogeneous costs for optimizing a European power system and for investigating the benefits from increased continent-wide transmission capacity limits. \citet{schlott2018impact} investigated the impact of climate change on a similar system with the same homogeneous cost assumption. Bearing the results of \citet{noothout2016diacore} in mind, the assumptions made in these studies appear questionable.

Besides the capital cost of generation assets, power system optimization models require the specification of several additional parameters, such as the electrical load at each node and each time step, the electrical properties of the transmission lines or the availability of the volatile renewable resources. Each of these parameters can only be specified within a certain range of uncertainty. Although recommended by \citet{decarolis2017formalizing}, these uncertainties are, however, only rarely addressed in current power system models \cite{mavromatidis2018uncertainty}. A possible way to characterize them has for instance been described by \citet{moret2017characterization}. Accordingly, variation ranges can be defined for uncertain parameters by either using values proposed in literature, modeling the variation or using historic data sets (e.g.).

The aim of this study is to investigate the effect of regional differences in cost of capital in Europe on the cost-optimal design of power systems with ambitious CO$_2$ reduction targets. Unlike other studies in the same field of research of recent years, we directly consider regional differences in the cost of capital. We investigate changes in expenditures for investment and operation compared to a homogeneous reference setup and relate these changes to differences in the cost-optimal deployment of generation capacity. Furthermore, we investigate the impact on overall system costs and the effect of diverging cost of capital. By doing so, our work contributes to a deeper understanding of the effect of input parameter uncertainties on the results of power system optimization models.

%----------------------------------------------------------------------------------------------
\section{Data and Methodology}
%-----------------------------------------------------
 \subsection{Power system expansion modeling}
 \label{sec:methodology}
%-----------------------------------------------------

In this paper, we investigate the impact of regional differences of the cost of capital on the cost-optimal design of a simplified highly renewable European power system. This cost-optimal design is derived from a greenfield expansion model. No existing generation, transmission and storage assets or retirement of these assets is considered. Instead, the cost-optimal expansion of all assets (starting from scratch) is determined using mathematical programming (Eq. (\ref{eq:minimisation}) - (\ref{eq:co2cap})). The power system considered comprises one node per country (30 in total) and 52 simplified transmission links connecting them. Electricity generation may stem from solar PV, on- and offshore wind , open cycle gas turbines (OCGT), run-of-river power plants and hydro reservoirs. Furthermore, we considered pumped hydro storage units and two generic storage types with fixed power-to-energy ratios (Table \ref{tab:costsassumptions}). Generation from coal and nuclear power plants has not been included for the following reasons: Firing coal for the generation of electricity contradicts the European Union's goals to significantly reduce greenhouse gas emissions. This goal is implemented by a global cap on CO$_2$ emissions (see below). Nuclear power is expected not to be cost-competitive. According to \citet{DIW_nuclear}, investing in nuclear power is uneconomical independent of the cost of capital, electricity prices and specific investment costs. \citet{lazard} found levelized costs for electricity (LCOE) for nuclear power to be between 112 and 183~USD/MWh, which is far above the LCOE for utility scale PV (43-53~USD/MWh) and onshore wind (30-60~USD/MWh). Furthermore, nuclear power exhibits a similar ratio between capital and marginal costs as the renewable resources, i.e. high upfront investments and almost no marginal generation costs. Thus, results likely hold if new developments render nuclear power a cost-efficient alternative.

In our model, the generation capacity of PV, wind and OCGT as well as the transmission capacity were expandable, but we fixed the capacity of hydro dams, run-of-river plants and pumped hydro storage units to the values published by \citet{kies2016restore}.

Different formulations of power system expansion models exist. The main difference is whether unit commitment of the generators or discrete expansion steps for the generator's nominal power or the transmission capacity are considered. In both cases, the optimization problem to be solved would be a mixed-integer problem. For this study, we use the pure linear approach, which has been used in several studies before, e.g. \cite{brown2018synergies, schlachtberger2017benefits, schlott2018impact}. The optimization problem used to derive the optimal system design contains investments in generation, storage and transmission capacity as well as hourly operational costs originating from load dispatch. It can be represented by:

\begin{align}
 \min_{g,G,f,F} &\sum_{n,s} c_{n,s} \cdot G_{n,s}  + \sum_{l} c_l \cdot L_l \cdot F_l + \sum_{n,s,t} o_{n,s} \cdot g_{n,s,t} \label{eq:minimisation}\\
\text{subject to} &\sum_s g_{n,s,t} - d_{n,t} = \sum_l K_{n,l} \cdot f_l \quad : \lambda_{n,t} \label{eq:powerbalance}\\
&{g}^-_{n,s,t} G_{n,s} \leq g_{n,s,t} \leq \bar{g}_{n,s,t} \cdot G_{n,s} \quad ,\forall n,t \label{eq:dispatchconstraint}\\
&\mathrm{soc}_{n,s,t} = (1-\eta_{n,s}^l) \cdot \mathrm{soc}_{n,s,t-1} + \eta_{n,s}^u \mathrm{uptake}_{n,s,t} ,\forall n, s, t > 1 \label{eq:storage_cont}\\
&\mathrm{soc}_{n,s,0} = \mathrm{soc}_{n,s,|t|} \quad ,\forall n, s \label{eq:storage_cycl}\\
&0 \le \mathrm{soc}_{n,s,t} \le \tau_{n,s} \cdot G_{n,s} \label{eq:soc_bounds}\\
%&f_{l,t} = \frac{\theta_{i,t} - \theta_{j,t}}{x_l} \quad ,\forall l,t \label{eq:pf}\\
&|f_l\left(t\right)| \leq F_l \quad ,\forall l \label{eq:transm_constraints}\\
%&G_{n,s}^\mathrm{min} \leq G_{n,s} \leq G_{n,s}^\mathrm{max} \label{eq:capacity_constraints}\\
%&F_l^\text{min} \leq F_l \leq F_l^\text{max} \label{eq:transm_cap_constraints}\\
&\sum_l F_l \cdot L_l \leq \mathrm{CAP}_{F} \quad : \lambda_{n,t}^{\textrm{trans}} \label{eq:global_transm_constr}\\
&\sum_{n,s,t} \frac{1}{\eta_{n,s}} \cdot g_{n,s,t} \cdot e_{n,s} \leq \mathrm{CAP}_{\text{CO}_2} \quad : \lambda_{n,t}^{\textrm{CO}_2} \label{eq:co2cap}
\end{align}
For an explanation of the used symbols see the nomenclature (Tab. \ref{tab:nomenclature} in the supplementary material). Constraint (\ref{eq:powerbalance}) describes the balance between generation and demand. Constraints (\ref{eq:dispatchconstraint}) - (\ref{eq:transm_constraints}) effect the dispatch and state of charge of generators, storage and transmission. The dispatch is constrained by the capacity -- or nominal power -- of the respective generator and/or storage unit (\ref{eq:dispatchconstraint}). 
In the case of storage units, the lower bound can be negative, i.e. when the storage takes up energy. In the case of generation technologies, the lower boundary equals zero. The potential generation $\bar{g}_{n,s}(t)$ describes the resource availability in case of fluctuating renewable generation facilities.
Constraints (\ref{eq:storage_cont}) and (\ref{eq:storage_cycl}) ensure storage consistency and cyclic usage of storage, i.e., state of charge at the beginning equals state of charge at the end of the investigated period. Constraint (\ref{eq:soc_bounds}) defines the bounds for the storage unit's state of charge. In equation (\ref{eq:storage_cont}) uptake$_{n,s,t}$ refers to the net energy uptake of the storage unit given by
\begin{equation*}
    \mathrm{uptake}_{n,s,t} = \eta_1 \cdot g_{n,s,t,\textrm{store}}  -  \eta_2^{-1} \cdot g_{n,s,t,\textrm{dispatch}}
    + \mathrm{inflow}_{n,s,t} - \mathrm{spillage}_{n,s,t}
\end{equation*}
where $\eta_{1,2}$ denote the efficiencies for storing and dispatching electricity, respectively. $g_{n,s,t,\textrm{store}}$ denotes the storing of electricity into the storage unit, $g_{n,s,t,\textrm{dispatch}}$ the dispatch. $\mathrm{inflow}_{n,s,t}$ is the natural inflow into the water reservoir of dams. And $\mathrm{spillage}_{n,s,t}$ denotes the amout of the natural inflow, which is spilled.
In addition, global limits on transmission and CO$_2$ emissions are enforced (Eq. \ref{eq:global_transm_constr} and \ref{eq:co2cap}, respectively). For this paper, we assumed a global limit of three times today's net transfer capacities ($3 \cdot 31.25~\text{TWkm}$) as an appropriate compromise between cost-optimal extension and technical and social concerns. Although, this assumption is slightly more conservative than the compromise grid defined in \citet{schlachtberger2017benefits} and \citet{brown2018synergies} at four times today's values, it allows to capture large parts of the benefits of distributing electricity from renewable resources due to the non-linear decrease in system costs with increasing transmission capacity \cite{schlachtberger2017benefits}. Inline with European emission reduction targets, we define a global CO$_2$ cap of 5\% of the historic level of 1990. In Eq. (\ref{eq:co2cap}), $e_{n,s}$ refers to the emissions given in tonnes of CO$_2$ equivalent per MWh of primary energy. $\eta_{n,s}$ denotes the efficiency of transforming one unit of primary energy into one unit of electrical energy ($g_{n,s,t}$). In this model, OCGT is the only generation type with non-zero CO$_2$ emissions. $e_{n,s}$ and $\eta_{n,s}$ are set to 0.18 tonnes per MWh and 0.39, respectively.

%Constraint (\ref{eq:pf}) describes the linearized power flow respected in the optimization and (\ref{eq:transm_constraints}) limits transmission along the single lines in order to account for thermal limits of short lines or voltage drops. To ensure n-1 security, the flows are additionally restricted to 70~\% of the transmission capacity, as it is commonly done in linearized power flow models. Since we limit the overall transmission capacity in (\ref{eq:global_transm_constr}), capital costs for transmission lines are set to zero and the second term in the objective function vanishes. As the LMPE and the CO$_2$ price, the price for transmission expansion is a result of our simulations (see Sec. \ref{sec:LMPE}).

The same methodology has been used to study, for instance, the impact of climate change  \cite{schlott2018impact}, synergies between sector coupling and transmission \cite{brown2018synergies}, the benefit of cooperation in a highly renewable European power system \cite{schlachtberger2017benefits} or the impact of CO$_2$ constraints \cite{weber2019counter}.
We used the software-toolbox \emph{Python for Power System Analysis} \cite{brown2017pypsa} and the commercial Gurobi solver to solve the optimization problem.

%----------------------------------------------------------------------------------------------
\subsubsection{Locational marginal prices for electricity}
\label{sec:LMPE}
%----------------------------------------------------------------------------------------------
Besides the cost-optimal deployment of generation capacity and the dispatch of electricity, the power system expansion modeling also delivers some information about how prices must develop to achieve this optimal system design via its \emph{dual problem}. For instance, if a global, i.e. system-wide, CO$_2$ cap constraint is enforced, the corresponding dual variable $\lambda^{\textrm{CO}_2}$ to this constraint yields a price for CO$_2$ emissions. If this price was set to the emissions in the primal problem without CO$_2$ constraint, the exact same optimization result would be obtained. In the same manner, the dual problem determines the price for electricity for each time step and at each bus within the system -- usually referred to as \emph{locational marginal price for electricity} (LMPE).

The dual problem corresponding to Eq. \ref{eq:minimisation}, again, is a linear problem. By applying the Karush Kuhn Tucker conditions, its objective function reads:
\begin{equation}
    \max_{\lambda,\lambda^{\textrm{CO}_2},\lambda^{\textrm{trans}}} \sum_{n,t}\lambda_{n,t}d_{n,t} + \lambda^{\textrm{CO}_2}\sum_{n,s,t}\frac{1}{\eta_{n,s}} g_{n,s,t} e_{n,s} 
    + \lambda^{\textrm{trans}} \sum_l F_lL_l \label{eq:dual_lp}
\end{equation}
At the global optimum, primal and dual problem are in equilibrium, Eq. (\ref{eq:minimisation}) equals Eq. (\ref{eq:dual_lp}), which is equivalent to:
\begin{equation}
    \sum_{n,s} c_{n,s} G_{n,s} + \sum_{n,s,t} o'_{n,s} g_{n,s,t} + \sum_l (c_l - \lambda^{\textrm{trans}}) L_l F_l = \sum_{n,t}\lambda_{n,t}d_{n,t} \label{eq:equilibrium}
\end{equation}
 with $\lambda^{\textrm{CO}_2}$ and $\lambda^{\textrm{trans}}$ being the global CO$_2$ price and the global price for expanding the transmission lines by one MWkm, respectively. Note that relaxing constraints (\ref{eq:global_transm_constr}) and (\ref{eq:co2cap}) leads to a decrease of the objective function. Hence, $\lambda^{\textrm{trans}}$ and $\lambda^{\textrm{CO}_2}$ are less than or equal to zero. $\lambda_{n,t}$ is the LMPE, which is the price consumers at a node $n$ and at time $t$ would have to pay for their electricity demand $d_{n,t}$ in equilibrium. According to Eq. (\ref{eq:equilibrium}), these payments cover the following three cost terms: (i) the regional investment in generation and/or storage capacity, (ii) the operational costs for generating electricity locally depending on the respectively available generation sources and (iii) the investment in transmission capacity. Here, the updated operational costs $o'_{n,s,t} = \left(o_{n,s} - \lambda^{\textrm{CO}_2}\nicefrac{e_{n,s}}{\eta_{n,s}}\right)$  also include the costs for CO$_2$ emissions and the effective costs for transmission reinforcements are given by the sum of the capital costs and the shadow price for transmission related to the global transmission capacity limit (Eq. \ref{eq:transm_constraints}).

%----------------------------------------------------------------------------------------------
\subsection{Cost of Capital Scenarios}
\label{sec:WACC}
%----------------------------------------------------------------------------------------------
  \begin{figure}
    \centering
    \includegraphics[width=.48\textwidth]{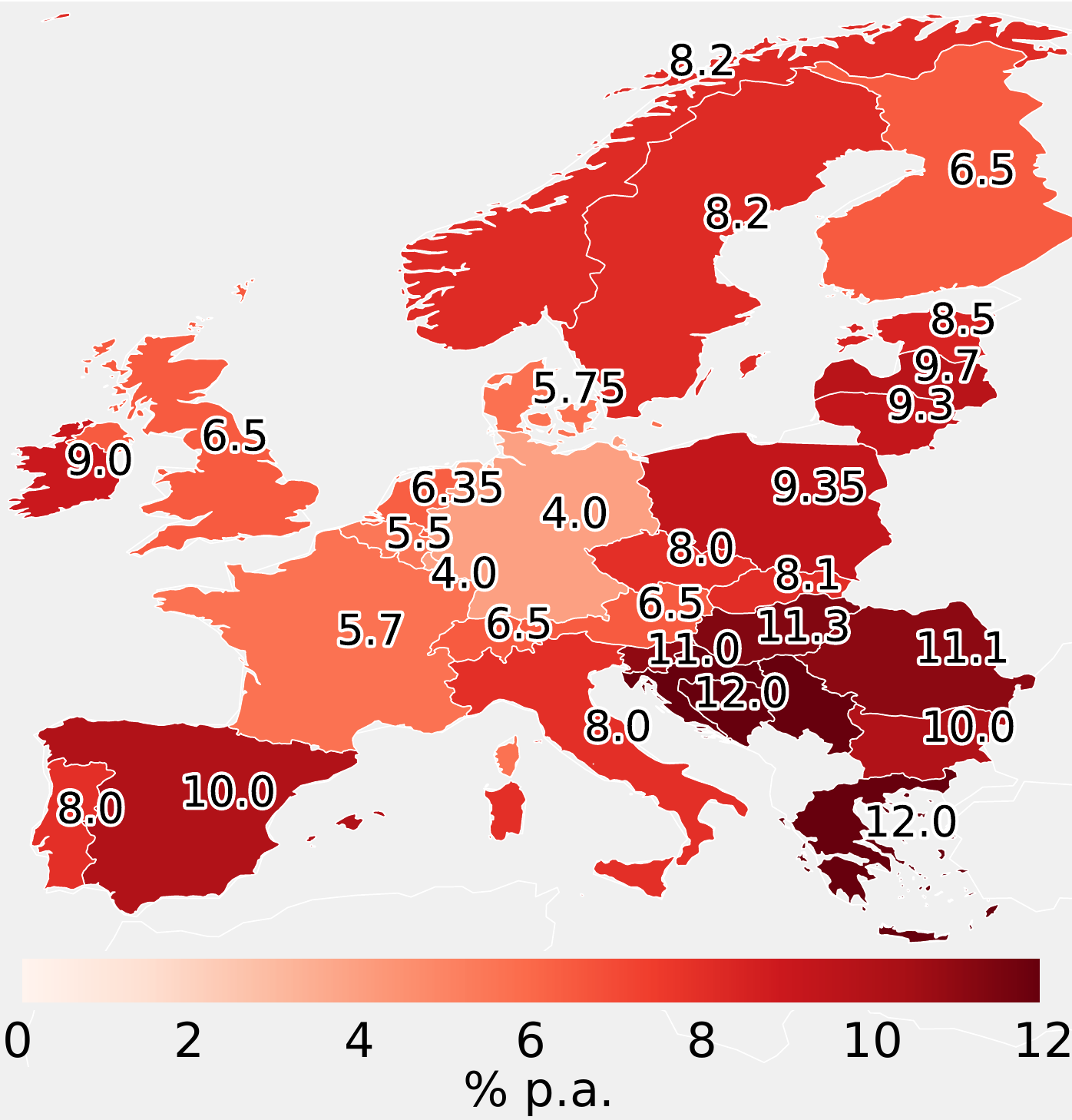}
    \caption{Weighted average cost of capital taken from \citet{noothout2016diacore}.}
    \label{fig:wacc_per_country}
\end{figure}
Investors raise funds from different funding sources and at different rates. The cost of capital -- when measured by the WACC -- is the weighted average of these rates. It comprises the costs of equity (between 6~\% and 15~\%) and debt (between 1.8~\% and 12.6~\%). If financial risk was equal across different regions, e.g by removing information asymmetry and taxes, the cost of equity and debt should be equal as well \cite{modigliani1958cost,modigliani1963corporate}. Hence, the WACC reflects varying market conditions between regions \cite{noothout2016diacore}. In order to investigate the impact of these varying conditions, we used the WACC obtained from \citet{noothout2016diacore} for the EU states. The WACC of the four remaining states in the area of interest have been assigned manually from their respective \emph{most similar} neighbouring country. Resulting WACC are shown in Fig. \ref{fig:wacc_per_country}. Germany has by far the lowest WACC at 4~\%, whereas the value peaks in South-Eastern Europe at 12~\%. In \citet{noothout2016diacore}, the WACC are only given for wind onshore projects, but reasons given for varying WACC  -- such as financing environment and policy risks -- hold true for all investments in renewables. We therefore assumed one uniform WACC for all generation and storage technologies per country. Because tariff-related risks are the major cause for discrepancies in WACC between European countries (\cite{temperton2016reducing}), it is unlikely that increasing shares of foreign direct investment or increasing capital investments due to increasing shares of renewables would have a relevant effect on WACC.
 
For this study, we varied the WACC for a number of scenarios:
\begin{enumerate}
    \item The \emph{today} scenario used the WACC values shown in Fig. \ref{fig:wacc_per_country}.
    \item The \emph{homogeneous} scenario considered a constant WACC of 7.1\% across Europe. 7.1\% was obtained as the demand-weighted average of WACC from the today scenario. The homogeneous scenario is used as the main reference throughout this study.
    \item For the \emph{inhomogeneous scenario}, the difference between every region and the average WACC has been doubled compared to the today scenario.
    \item For nineteen scenarios indexed with the numbers -9 to 9, the WACC was linearly interpolated between the homogeneous scenario (index 9) and the inhomogeneous scenario (index -9) to investigate potential path-dependencies.
 \end{enumerate}

%----------------------------------------------------------------------------------------------
%\subsubsection{Cost assumptions}
%\label{sec:costs}
%----------------------------------------------------------------------------------------------
%Throughout this study, several cost terms are used: The \emph{cost of capital} refers to the rate, which an investor is expected to pay on average to finance its assets. As \citet{noothout2016diacore}, we measure it via the WACC. 
When used as the return rate $r$, the WACC determines the annuity $a_{n,s}$ of an asset $s$ in region $n$ with lifetime $l_s$ via
\begin{equation}
    a_{n,s} = \frac{r_n \left(1+r_n\right)^{l_s}}{\left(1+r_n\right)^{l_s}-1}    
\end{equation}
From these annuities and the \emph{investment costs} $c^\textrm{inv}_{n,s}$ the annualized \emph{capital costs} $c_{n,s}$ of an asset can be computed via
\begin{align}\label{eq:WACC}
    c_{n,s} &= c^\text{inv}_{n,s} \cdot a_{n,s}
\end{align}
They are given in Euro per MW per year for a generator $s$ in region $n$.

The \emph{investment} refers to the product of the capital costs and the installed capacity $G_{n,s}$. Similarly, the \emph{operational costs} are defined as the product of the \emph{marginal costs}, i.e. the costs for generating one (additional) unit of electricity, of a generator $o_{n,s}$ and the actual generation $g_{n,s,t}$. Capital costs $c_{n,s}$ and operational costs $o_{n,s}$ for all technologies are given in Table \ref{tab:costsassumptions}. 

We derived a cost-optimal design of a European power system with an ambitious CO$_2$ reduction target by solving the previously introduced optimization problem for the four aforementioned scenarios. Regional expenditures for investment and operation were measured as:
\begin{equation}
    \begin{aligned}
        I_n &= \frac{\sum_s c_{n,s} \cdot G^*_{n,s}}{\sum_t d_{n,t}} \\
        &\text{and} \\
        O_n &= \frac{\sum_{s,t} o'_{n,s} \cdot g^*_{n,s,t}}{\sum_t d_{n,t}}
    \end{aligned}
    \label{eq:levelized}
\end{equation}
respectively and compared between the simulations. Here, the asterisk indicates the optimal solution obtained from solving Eq. (\ref{eq:minimisation}).

In Eq. (\ref{eq:minimisation}) through (\ref{eq:co2cap}), investment in the inter-connecting transmission grid is included via cost-optimization as well as through the assumption of a global limit on transmission capacity extension. This reflects that the expansion of the power system often is not purely a technical-economical problem. Instead, it is often hampered by political and social constraints -- such as missing public acceptance, for instance -- meaning that although transmission grid extension might be cost-optimal, it cannot be realized. Consequently, the costs for transmission expansion can only be estimated with great uncertainty. As shown above one can, nevertheless, derive a shadow price for transmission grid expansion $\lambda^\text{trans}$. However, this shadow price is a \emph{political} price, which cannot directly be compared to the market prices $c_{n,s}$ and $o_{n,s}$. Therefore, we focus on the investment in generation and storage assets in Sec. \ref{sec:results}.

\begin{table*}%[!ht]
    \begin{center}
        \caption{Cost assumptions for generation and storage technologies, originally based on estimates from \citet{schroder2013current} for the year 2030; fixed operational costs are included in the capital costs.}
        \resizebox{\textwidth}{!}{\begin{tabular}{ lrrrrrr }
        \hline
            Technology & Capital Cost & Marginal cost & Efficiency & Lifetime & energy-to-power ratio \\ 
               &  [EUR/GW/a] & [EUR/MWh] & dispatch/store & [years] & [hours]\\ 
            \hline
            OCGT & 47,235 & 58.385 & 0.390 & 30 &\\
            Onshore Wind & 136,428 & 0.015 & 1 & 25 &\\
            Offshore Wind & 295,041 & 0.020 & 1 & 25 &\\
            PV & 76,486 & 0.010 & 1 & 25 &\\
            Run-Off-River & -- & 0 & 1 & -- &\\
            Hydro Reservoir & -- & 0 & 1 / 1 & -- &\\
            PHS & -- & 0 & 0,866 / 0,866 & -- &\\
            Hydrogen Storage & 195,363 & 0 & 0.580 / 0.750 & 30 & 168\\
            Battery & 120,389  & 0 & 0.900 / 0.900 & 20 & 6\\
            \hline
        \end{tabular}}
        \label{tab:costsassumptions}
    \end{center}
\end{table*}

%----------------------------------------------------------------------------------------------
\subsubsection{Generation and load data}
%----------------------------------------------------------------------------------------------
We use one year of hourly availability data for onshore wind, offshore wind and solar PV as described by \citet{kies2016restore}. The underlying weather data stems from the MERRA reanalysis \cite{Merra} as well as Meteosat First and Second Generation. Feed-in from wind has been modeled using the power curve of an Enercon E-126 at $140$ m hub height. AC power from PV modules has been simulated by applying the Heliosat method \cite{Cano_1986, Hammer_1998}, the Klucher model \cite{Klucher_1979} as well as the parameters of a Sunny Mini Central 8000TL converter.
%To model the distribution of wind and PV generation facilities within each country, the German distribution of capacities for wind and PV in dependency of the available resource (average wind speed / average global horizontal irradiation) was empirically derived and adopted for every country. 
The natural inflow to hydro dams and run-of-river power plants is taken from \citet{kies_zenodo}. It has been modeled as a linear function of the potential energy of the the run-of data obtained from the ERA Interim reanalysis \cite{dee2011era}. More details on the methodology applied to generate the time series are given by \citet{kies2016restore, kies2016effect}.
%The potential gravitational energy of a mass $m$ relative to the sea level is given by
%\begin{align*}
% U &= m g h, 
%\end{align*}
%where $g = 9.81 \text{m}/\text{s}^2$ is the constant of gravititational acceleration on Earth and h the height above sea level.
%\begin{align*}
% I^H_n(t) &=  f \int_A g m(t) h(x,y) dA.
%\end{align*}
%$f$ is a normalization constant to enforce $\left<I^H_n(t)\right>$ = $\left<G(t)^H_n\right>$, where $G_n^H$ is today's average hourly generation from hydro in the corresponding country.
Load time series were derived from historical load data provided by ENTSO-E and modified within the RESTORE 2050 project to account for expected increasing shares of e-mobility and electric heat pumps \cite{restore}.

%-----------------------------------------------------
\section{Results}
\label{sec:results}
%-----------------------------------------------------
\begin{figure}
    \centering
    \includegraphics[width=.48\textwidth]{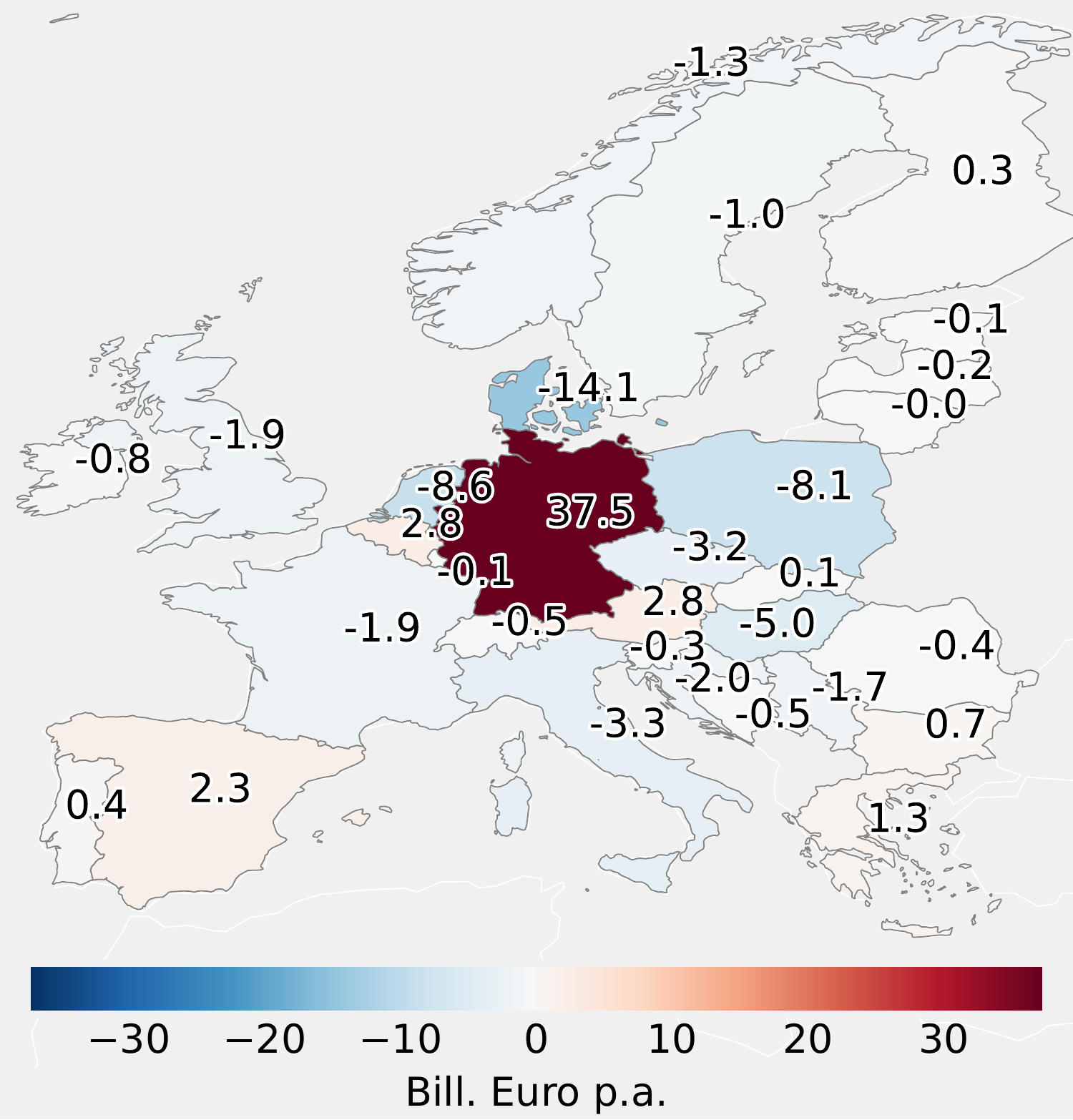} \\
    \includegraphics[width=.48\textwidth]{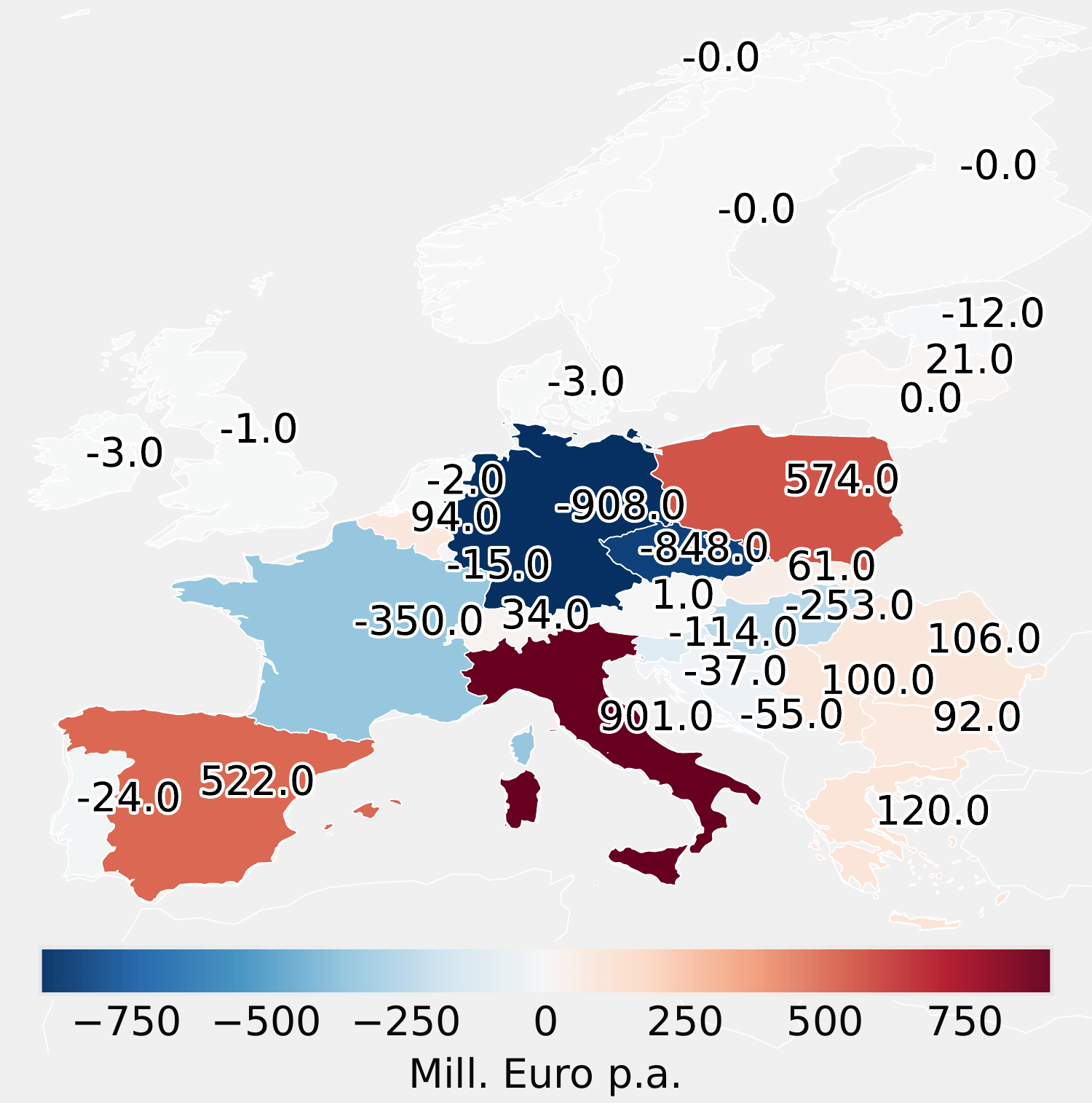}
    \caption{Change in regional investment in generation and storage units (top) and  operational costs (bottom) between the homogeneous and the today scenario.}
    \label{fig:diff_investments}
\end{figure}

\begin{figure}
    \centering
    \includegraphics[width=.48\textwidth]{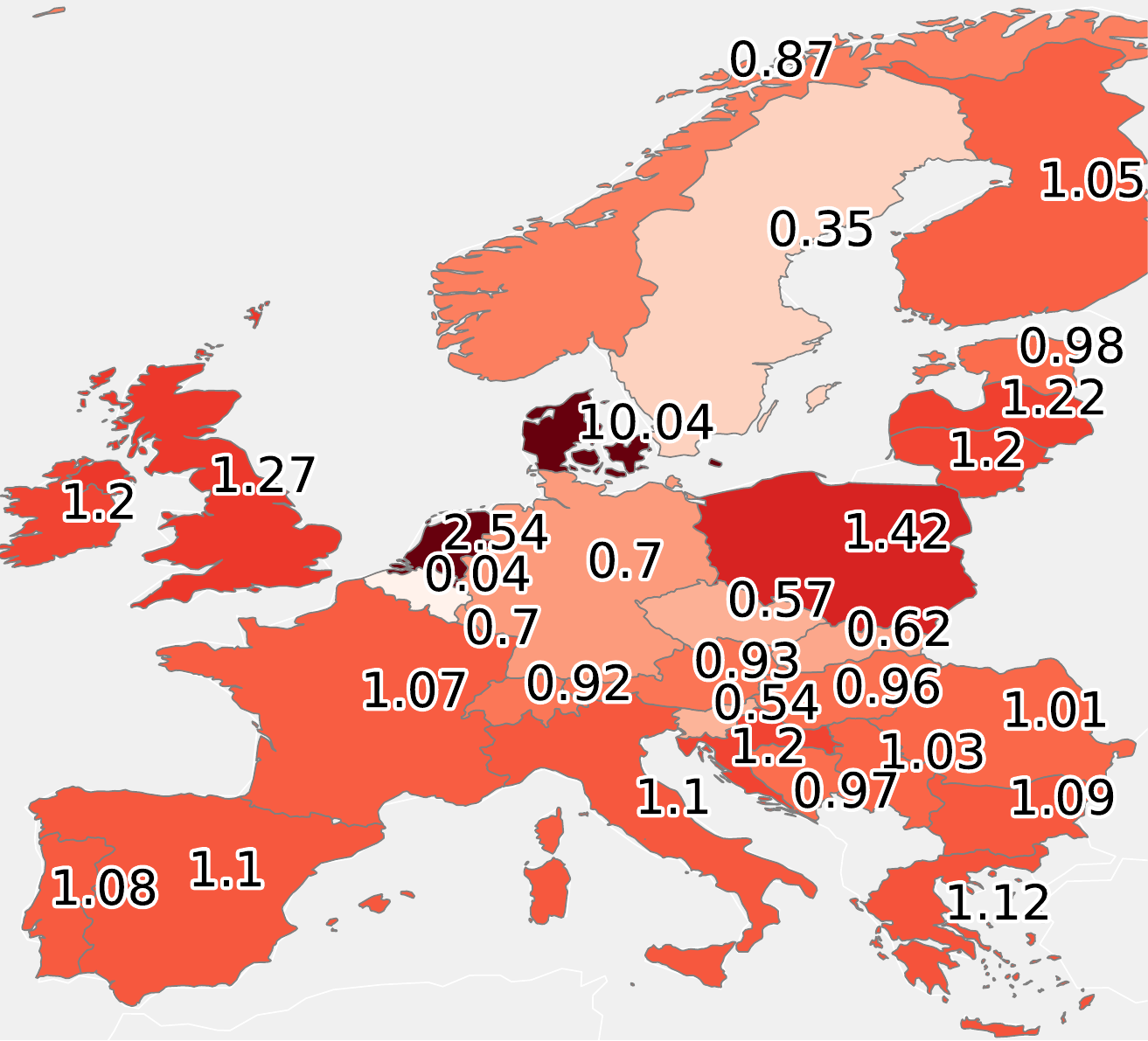} \\
    \includegraphics[width=.48\textwidth]{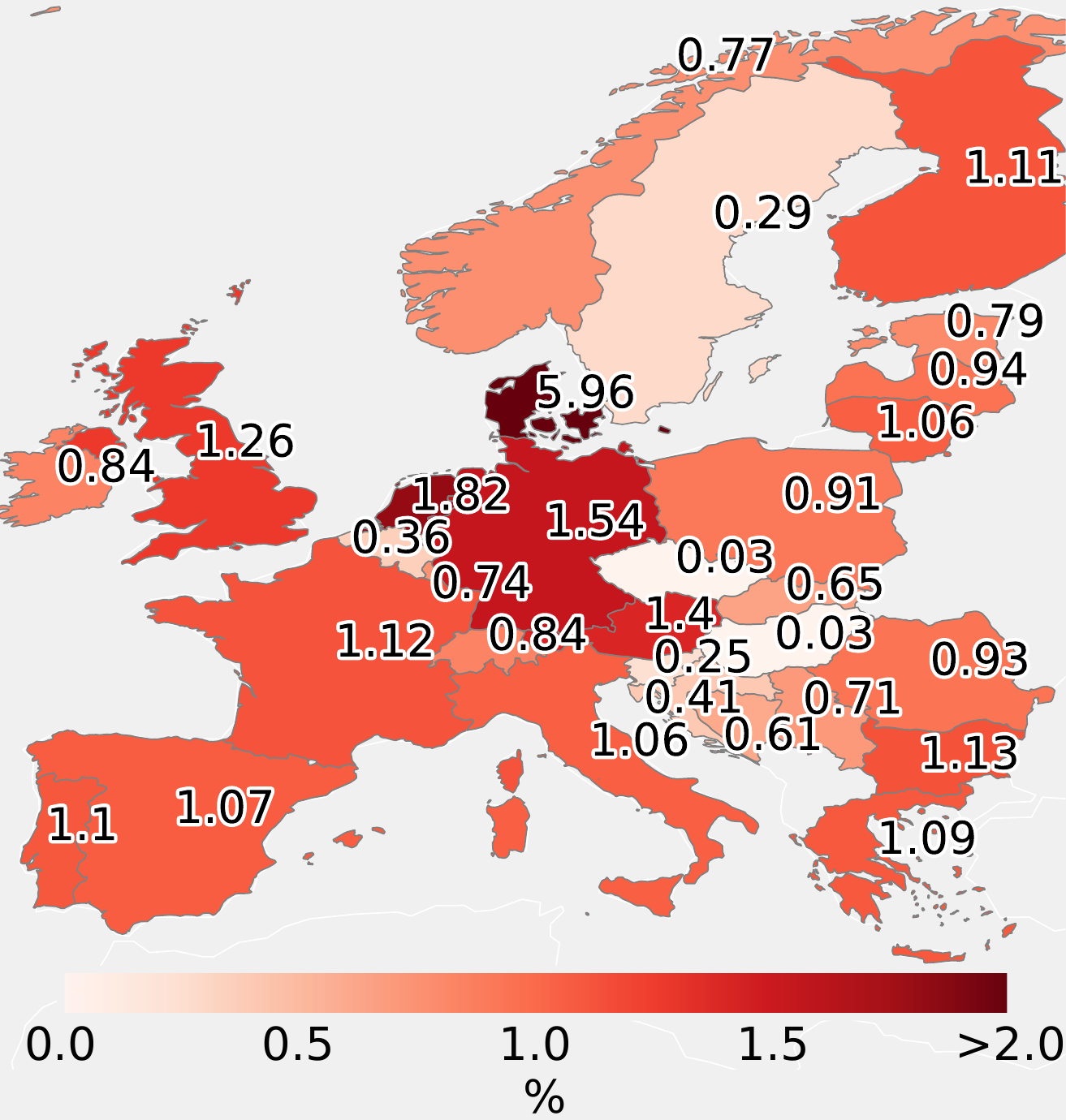}
    \caption{Electricity penetration (local generation divided by demand) in the homogeneous scenario (top) and the today scenario (bottom).}
    \label{fig:diff_penetration}
\end{figure}

If the regional distribution of the cost of capital would not affect the optimal system design, investments would change between these scenarios in the same way as there is a difference in the WACC: It would rise in countries where cost of capital increases and shrink where cost of capital decreases compared to the reference. Instead, we observe highest increases in investments in the country, which exhibits the lowest WACC: Germany. Here, investments in generation and storage assets increase by approximately 37.5 Billion Euro per annum or 96~\% (Fig. \ref{fig:diff_investments} top). Increasing investments are also observed for Belgium, Austria, Spain, Greece, Bulgaria, Portugal and Finland. Of these countries, Belgium, Austria and Finland exhibit relatively low cost of capital like Germany, while they are relatively high in the others (Fig. \ref{fig:wacc_per_country}). This evidences significant changes in system design, expressed by changes in optimal capacity deployment, which in turn influences regional operational costs (Fig. \ref{fig:diff_investments} bottom).

\begin{figure}[htb]
    \centering
    \includegraphics[width=.45\textwidth]{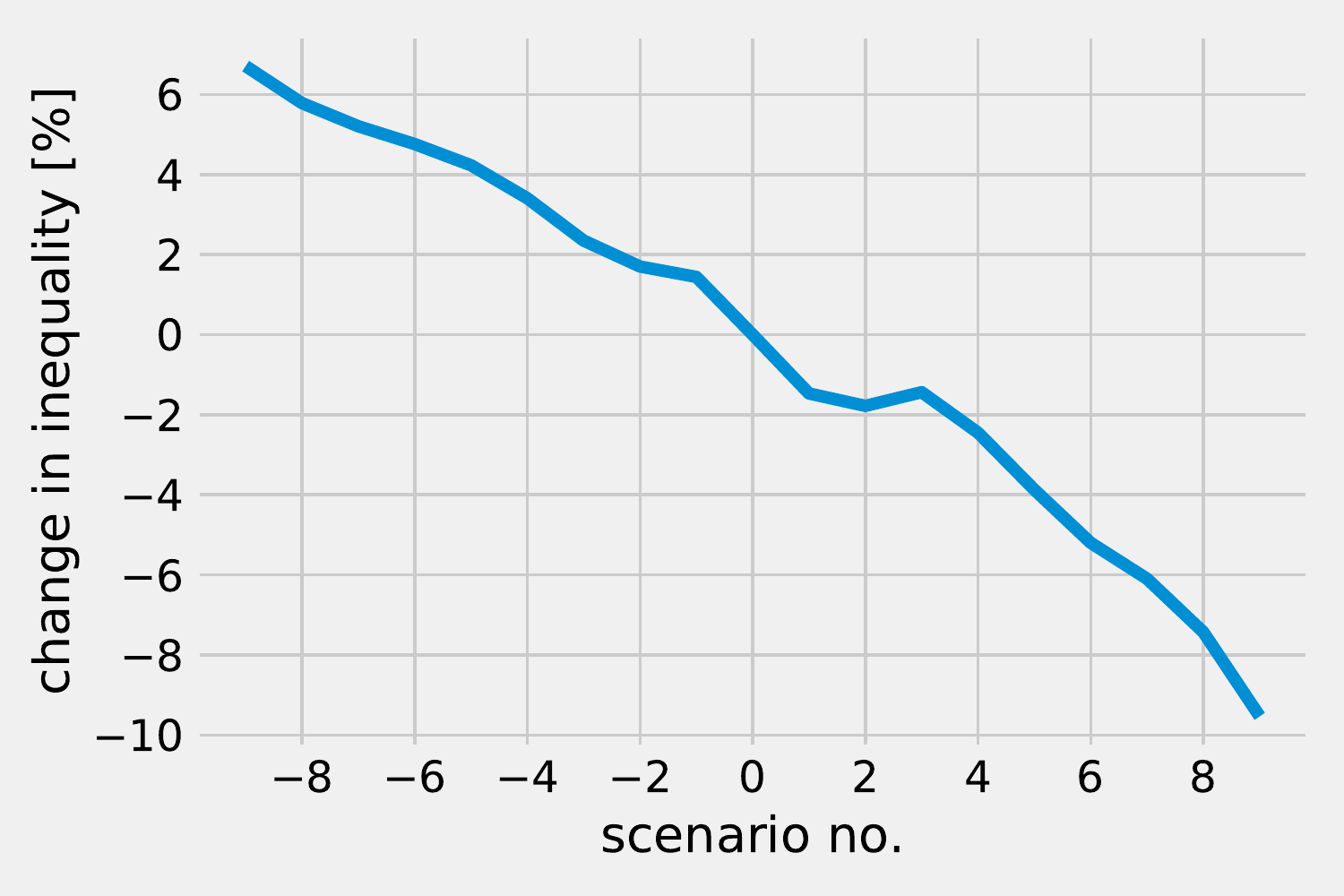}
    \includegraphics[width=.45\textwidth]{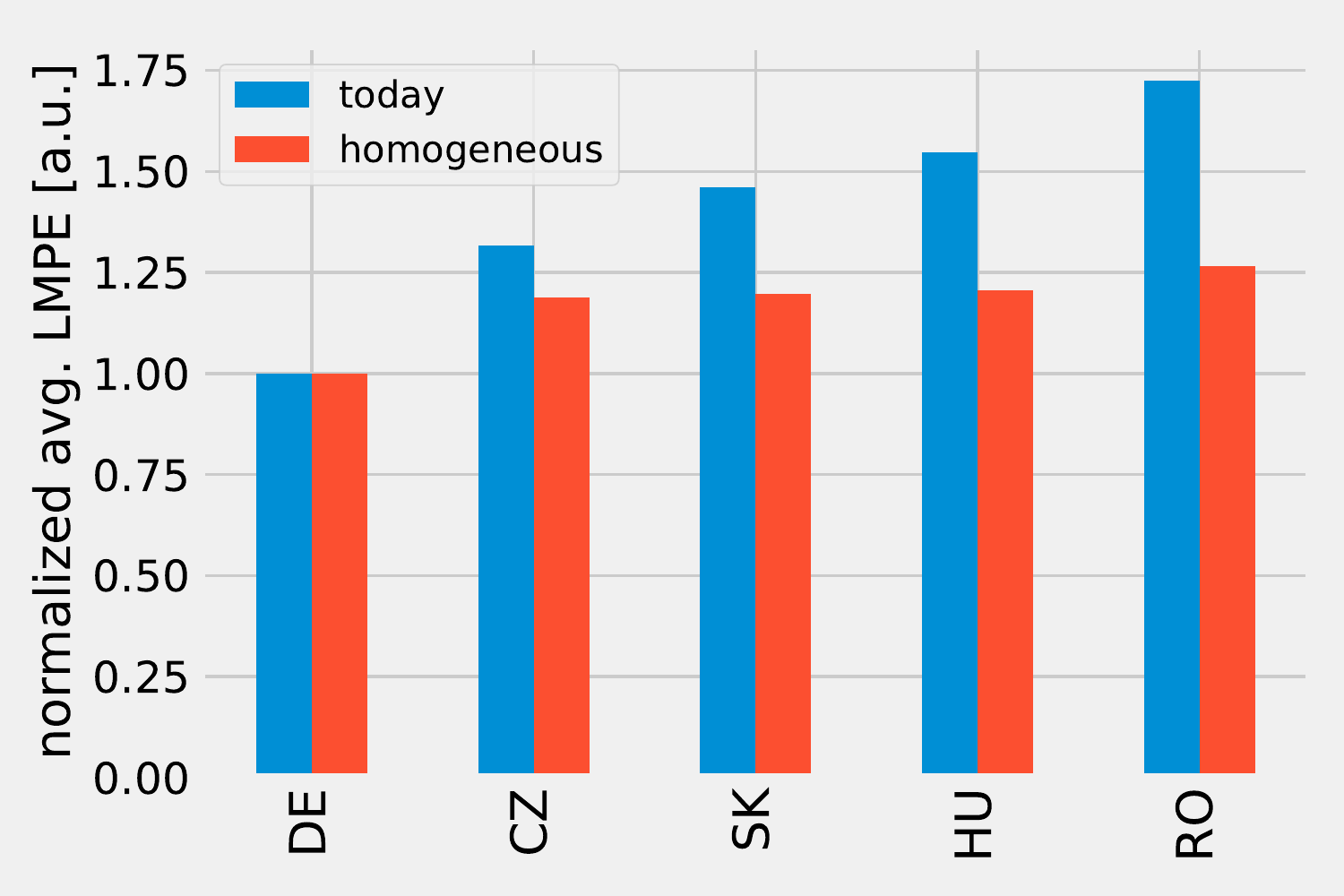}
    \caption{Difference in inequality in regional expenditures for electricity as measured by the 20:20 ratio [a.u.] relative to the today scenario (left) and cascade in LMPE for four countries downstream of Germany normalized to the LMPE of Germany [a.u.] (right).}
    \label{fig:20:20}
\end{figure}

Compared to the homogeneous scenario, inhomogeneous WACC lead to a strong agglomeration of power generation capacity in Central-Western Europe, especially in Germany, France, Austria and Belgium (see supplementary material \ref{sec:supplementary}). Consequently, the penetration rate, i.e. the ratio of local electricity generation over local demand, in Germany and Austria increases by 120~\% and 50~\% respectively (Fig. \ref{fig:diff_penetration}). While both are net importers in the homogeneous scenario, they become net exporters in the today scenario. In turn, many countries in Eastern Europe exhibit a higher dependency on imports in the inhomogeneous scenario. This is reflected in penetration rates below one and in a distinct step-wise increase in the locational marginal price for electricity the further the respective country is from the exporting countries in Central-Western-Europe (Fig. \ref{fig:20:20} right).

We assume equal marginal costs for each type of generator, no matter at which node the generator is located. Hence, a rise in operational costs as depicted in Fig. \ref{fig:diff_investments} (bottom) can only be caused by a replacement of generators with low marginal costs, i.e. wind and PV, with gas power plants and/or the intensified use of gas power. Again, this shift from one generation technology to another between the today and the homogeneous scenario is most pronounced in Germany. Expenditures for the regional operation of power plants decrease by approximately 908 Million Euro per annum. France and the Czech Republic profit from the cheap electricity generation in Germany (-350 and -848 Million Euro per annum, respectively). And Hungary increasingly imports electricity from Austria, which leads to a decrease in regional operational costs of 253 Million Euro per annum. In Italy, Poland, Spain and the South-Eastern European countries, however, regional expenditures for operation increase due to the intensified deployment and use of less capital-intensive gas power (see supplementary material \ref{sec:supplementary}). Overall, a slight increase in operational costs is observed due to an increased generation share of offshore wind power in the today scenario compared to the homogeneous scenario.

\begin{figure}[htb]
    \centering
    \includegraphics[width=.48\textwidth]{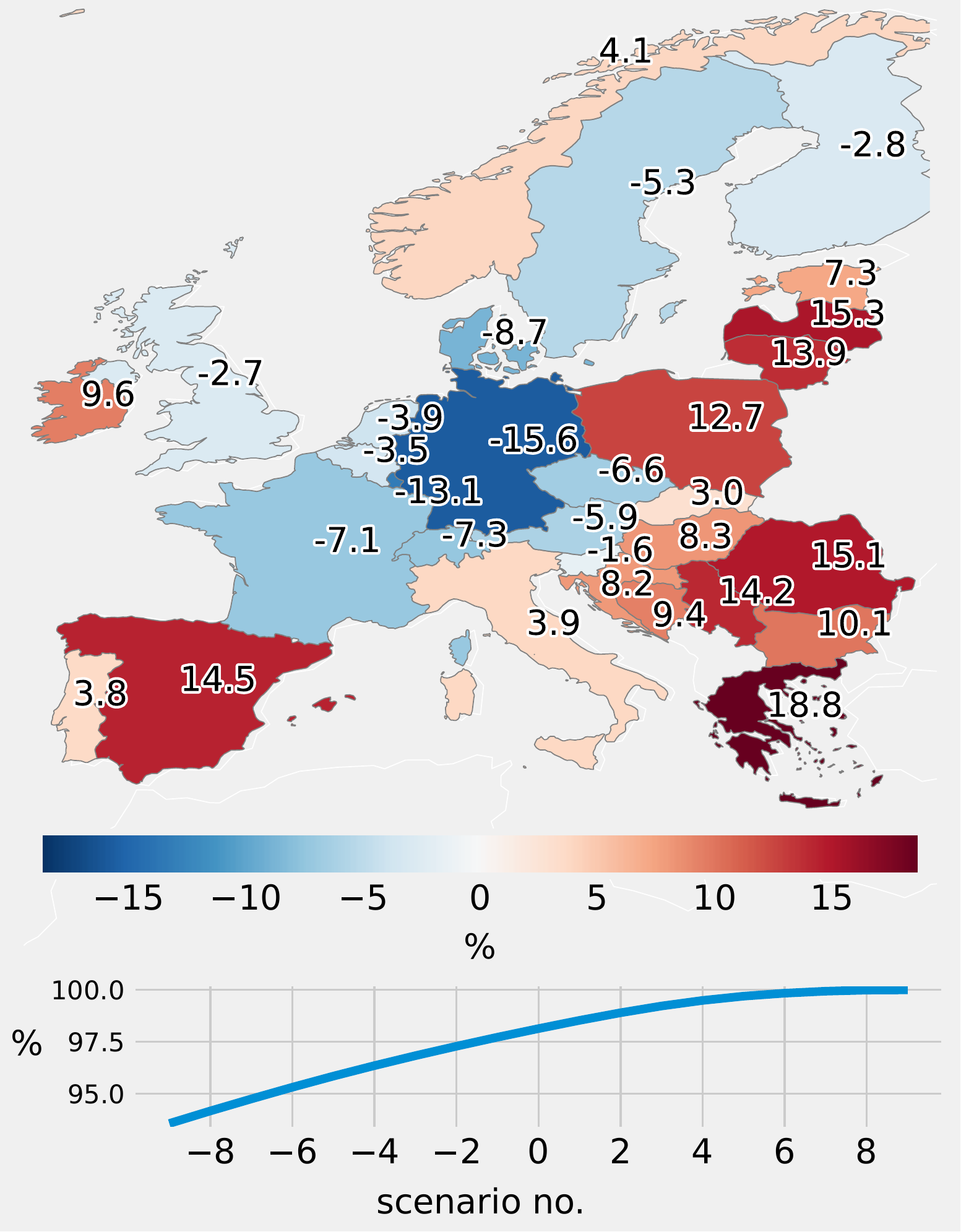}
    \caption{Regional relative change in locational marginal prices for electricity between the homogeneous scenario and the today scenario. The insert shows the system costs normalized to the systems costs of the homogeneous scenario.}
    \label{fig:expenditures_relchange}
\end{figure}

In combination, the mentioned effects have different implications for different countries in Europe concerning the optimal deployment of the different generation technologies and the regional expenditures. In Germany, for instance, below-average WACC would lead to increasing overall investment and capacity deployment, especially of capital intensive wind power (+280~GW, see supplementary material \ref{sec:supplementary}). Nevertheless, LMPE would fall due to significantly higher penetration rates (+120~\%, Fig. \ref{fig:diff_penetration}) and a lower dependency on gas power (-11~GW) and imports (Fig. \ref{fig:expenditures_relchange}). The Czech Republic profits from its proximity to the 'export country' Germany. The fact that Germany exports electricity reduces the need for local investments and the expenditures for operation in the neighbouring Czech Republic. Transmission cost are low, because the two countries are directly connected, and consequently the LMPE decreases. A similar effect can be observed in Sweden and Denmark. In Hungary, LMPE rise, although the expenditures for investment and operation and the penetration decrease here as well. This is due to the fact that Hungary is located further away from the exporting countries (Fig. \ref{fig:20:20} right). Thus, LMPE need to cover more investment in transmission capacity upstream.

Wind power in general -- both onshore and offshore -- profits from relatively low WACC in Germany as, to a smaller extent, in Belgium, France and Finland. Besides relatively low cost of capital, these countries exhibit relatively good wind resources and, at least in the case of Germany, high demand for electricity and a favourable topological position within the network. This leads to the aforementioned additional 280~GW onshore wind power in Germany, 24~GW in France, 7~GW in Belgium and 4~GW in Finland in the today scenario compared to the homogeneous reference. Additional offshore wind power deployment only occurs in Germany: +30~GW. In most other countries, onshore wind power deployment drops: Denmark -77~GW, Poland -65~GW, the Netherlands -47~GW, Italy -39~GW. In sum, continent-wide onshore and offshore wind power installations increase by 8~GW and 21~GW, respectively. Figures showing the regional changes in nominal power and investment per generation source can be found in the supplementary material \ref{sec:supplementary}.

In contrast to wind power, continent-wide solar PV installations decrease by 24~GW between the homogeneous scenario and the today scenario. The main reason for this is the reduction of deployment of PV power in countries with high cost of capital, especially the Czech Republic (-30~GW), Hungary (-22~GW), Croatia (-16~GW) and Serbia (-11~GW). As described above, this reduction is not driven by the costs of capital alone, but triggered by the relative proximity to countries with relatively low WACC, i.e. Germany and Austria. In these countries increased deployment of PV power can be observed (Germany +6~GW, Austria +38~GW). In Belgium PV power deployment rises by 26~GW. In Spain, PV replaces 10 GW of even more capital-expensive wind power.

It has been shown by \citet{tranberg2018flow} that solar PV and gas power are suitable complements. In countries with relatively good solar resources, gas power plants -- potentially together with battery storage units -- are used to cover load peaks and to bridge times of low sunshine. Consequently, as PV power deployment reduces, the continent-wide deployment of gas power plants decreases by 10~GW in the today scenario compared to the homogeneous scenario as well.

\begin{figure*}
    \centering
    \includegraphics[width=.48\textwidth]{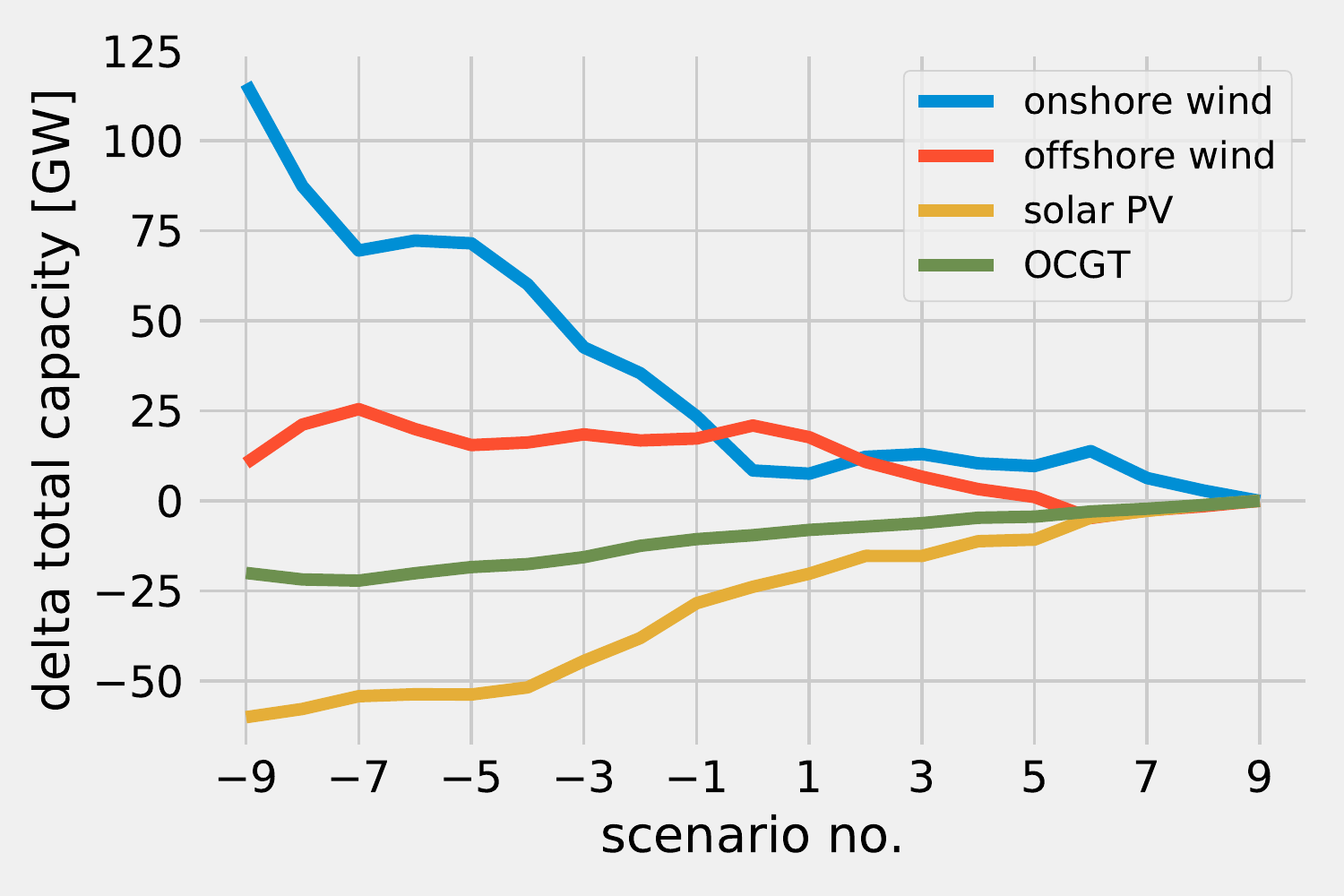}%
    \includegraphics[width=.48\textwidth]{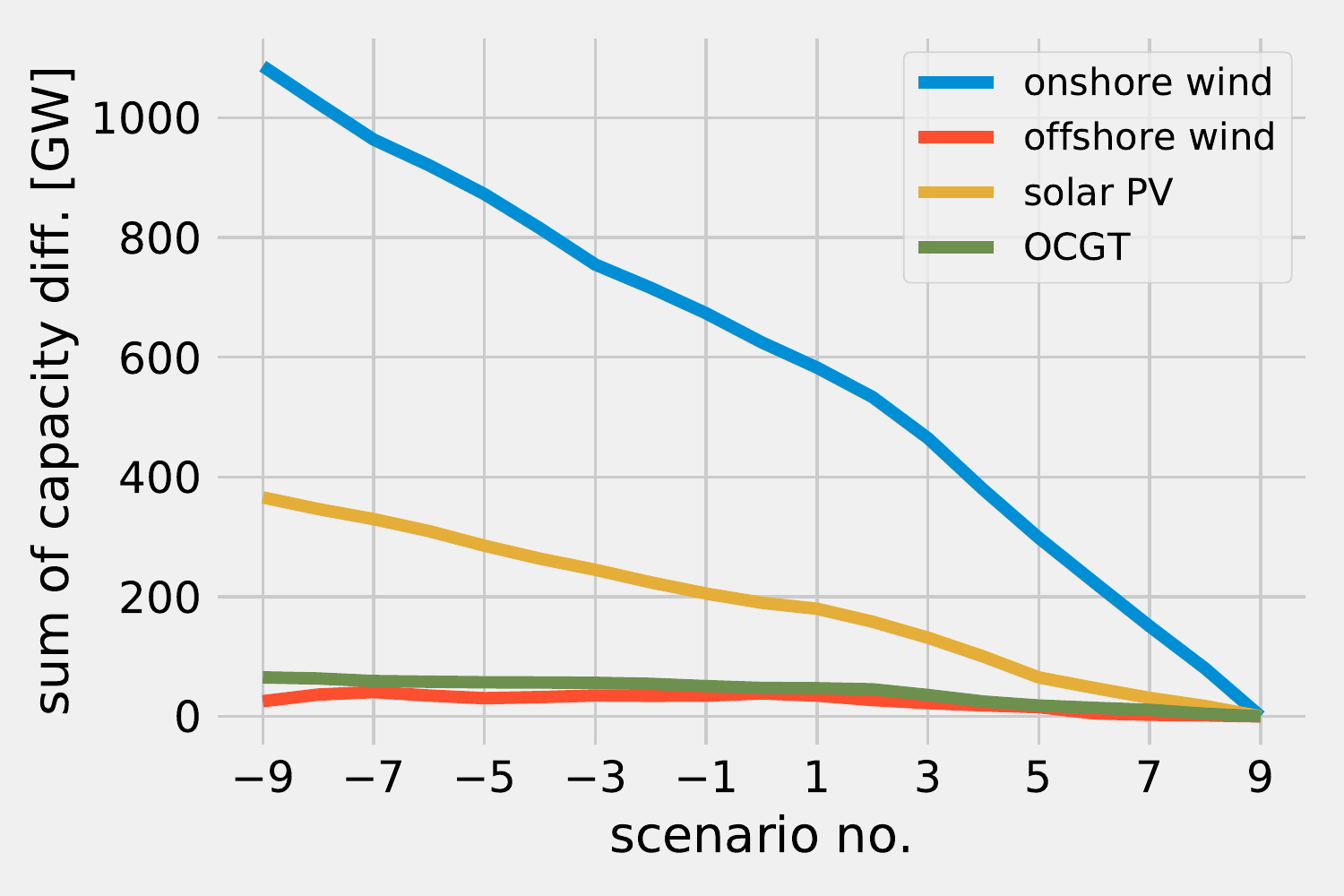}
    \caption{Left: Change in overall power generation capacity deployment for various generation sources; Right: Sum of absolute nodal differences in power generation capacity deployment for various generation sources.}
    \label{fig:delta}
\end{figure*}

It is also noteworthy that, although operational costs only decrease slightly, overall system costs significantly dependent on the spatial distribution of the WACC. Compared to the homogeneous scenario, inhomogeneous WACC lead to a reduction of levelized costs of electricity of approximately 2.5~\% in the today scenario and of up to more than 5~\% in the inhomogeneous scenario (insert of Fig. \ref{fig:expenditures_relchange}). This reduction in LCOE is due to the complex interaction of regional demand, costs, the quality of the volatile renewable resources and the capabilities of transmitting and storing electricity. We describe this complex interaction by investigating the levelized costs for electricity separately for each generator type. It is defined as the sum of generator type specific investment and operational costs divided by the electricity generation. Since the implemented CO$_2$ cap (Eq. (\ref{eq:co2cap})) is reached in each scenario, the generation from gas power plants -- the only generator type with non-zero CO$_2$ emissions considered -- does not change between the scenarios. Additionally, a decrease in gas power capacity installation can be observed (Fig. \ref{fig:delta} left). Hence, the utility rate for gas power increases in the today scenario compared to the homogeneous scenario. However, because gas power is increasingly deployed in countries with relatively high cost of capital, levelized cost for gas power only decrease slightly by less than 1~\%. A similar effect can be observed for solar PV: As mentioned above, PV power suffers from the relatively high cost for capital in the countries with relatively good solar resources. This leads to an increase in levelized costs for PV power of approximately 6~\%. Consequently, the decrease in LCOE must be driven by a reduction in the levelized costs for wind power. Indeed, the levelized cost for onshore wind power decreases by around 8~\% and for offshore wind power by 15~\% between the today and the homogeneous scenario caused by the co-occurrence of high demand, relatively low costs and good wind resource quality in the Central-Western European countries mentioned already. A slight increase in the levelized costs for storage units (batteries +1~\%, H$_2$ +2~\%) counteracts this effect. For storage units, operational costs have been determined from the product of LMPE and energy uptake as suggested by \citet{pawel2014cost}.
%It is to a large extent driven by the cost reduction in the high-demand countries in Central-Western Europe, especially Germany, France and the UK, due to the relatively low WACC in these countries. Together, the countries, which exhibit a below average WACC, make up approximately 56~\% of the system-wide electricity demand. The cost reduction in these countries and the affected neighbouring countries over-compensates the increase in costs in the remaining countries and, consequently, leads to the observed decrease in LCOE.

However, the reported reduction in costs is unevenly distributed among countries. In 17 out of 30 countries, LMPE rise due to a relative increase of cost of capital leading to higher local investments, higher shares of generators with low capital but high marginal costs and a higher dependency on imports, i.e. investment in transmission. This causes a growth in inequality of regional expenditures for electricity of up to 10~\% in the today scenario compared to the reference scenario (Fig. \ref{fig:20:20} left). Here, inequality is measured as the so-called \emph{20:20 ratio} defined as the ratio between the 20th and 80th percentile of the levelized nodal expenditures for electricity. One main reason for this rising inequality are higher LMPE in countries with weaker economies, e.g. Greece, Romania, Serbia, Latvia and Lithuania (Fig. \ref{fig:expenditures_relchange}). In these countries the Human Development Index is below the value of, for instance, Germany and France \cite{HDI2018}. Higher LMPE would put an additional burden on electricity consumers in these countries, which potentially suffer from economic hardship already and can, therefore, hamper the acceptance of renewables and the mitigation towards climate goals.

Overall, it has been demonstrated that diverging WACC lead to an increased inequality in regional expenditures for electricity and a reduction of levelized costs of electricity -- driven by changes in regional investment and resulting changes in regional generation and electricity penetration. %Detailed results concerning capacities, CO$_2$ prices and costs are shown in Table \ref{tab:key_findings_numbers}.
We also explained that wind power profits from low cost of capital in wind-rich, high-demand countries such as Germany, France and Belgium. If WACC diverge further, this effect accelerates, leading to additional 116~GW onshore wind power capacity for the inhomogeneous scenario compared to the homogeneous case (Fig. \ref{fig:delta} left). Offshore wind power, however, remains more or less stable around an additional 11~GW. The deployment of gas and PV capacity simultaneously falls. Besides these changes in overall capacity expansion, there is a significant redistribution of deployment between countries (right panel of Fig. \ref{fig:delta}). This redistribution is most pronounced for onshore wind power: Compared to the homogeneous scenario, more than 1~TW are shifted from one country to another, expressing the agglomeration of wind power in Germany, France and Belgium mentioned earlier. Additionally, the inequality in regional expenditures for electricity rises up to 16~\% in the inhomogeneous case compared to the homogeneous scenario (Fig. \ref{fig:20:20} left).

%-------------------------------------------------------------------------------
\section{Discussion}
\label{sec:Discussion}
%-------------------------------------------------------------------------------
Inline with \citet{hirth2016role} our study emphasizes the importance of cost of capital in the context of fostering the integration of renewable generation sources into future power systems. We show that regional differences in weighted average cost of capital lead to significant changes in the optimal design of a European power system with an ambitious CO$_2$ reduction target when compared to a setup with homogeneous WACC. The latter is an often made assumption in power system modeling and -- as we explained -- potentially leads to wrong conclusions concerning the optimal system design and the spatial distribution of costs. \citet{schlachtberger2017benefits} for instance assumed a constant return rate of 7~\% across Europe. They reported an optimal cost share for solar PV of at least 30~\% (depending on the global transmission capacity limit) in Hungary. Considering the relatively high WACC in Hungary we find an optimal share of solar PV of less than one per cent of the annualized system costs. Similarly, \citet{schlachtberger2017benefits} found a PV cost share of 40~\% to approximately 75~\% in Austria. In our study, Austria exhibits a relatively low WACC. Together with its position within the network and good solar resources this leads to a PV cost share of 91~\%. Furthermore, the assumption of homogeneous WACC across Europe leads to too conservative estimates of the levelized costs of electricity, mainly caused by too high estimates of the levelized costs for wind power.

The relative increase of wind power in Europe in the case of inhomogeneous WACC also modifies the variance of the generation time series. Wind power in general exhibits lower diurnal variability and a seasonal cycle opposite to the seasonal cycle of solar PV \cite{heide2010seasonal}. Hence, modifications of the generation mix lead to different requirements for flexibility as well. Consequently, our results -- at least partially -- change the interpretation of other studies investigating the optimal generation mix, the need for flexibility options and/or the interplay of generation and storage under the assumption of homogeneous cost of capital across Europe: e.g. \cite{brown2018synergies, schlott2018impact, tranberg2018flow}. \citet{schlott2018impact}, for instance, found an increasing importance of PV power in Europe under different climate change scenarios. This effect might be weakened when the relative high costs for investments in the Southern European countries are taken into account. \citet{brown2018synergies} describe how the integration of battery electric vehicles (BEV), long-term thermal energy storage (LTES) and power-to-gas units (P2G) helps smoothing the variability from solar and wind power generation in a sector-coupled European power system. In particular, BEV interact with the diurnal variability of solar power and load, while LTES and P2G mainly balance the synoptic to seasonal variations. These findings are confirmed by \citet{tranberg2018flow}. Since synoptic variations are more pronounced for wind power (for solar power the diurnal cycle dominates), an increased share of wind power might favour LTES and P2G while decreasing the importance of short-term storage, such as batteries and BEV.

Some critical remarks: Although our model setup is inline with a number of similar studies, the concept of foreign direct investment (FDI) is not considered. \citet{wall2019policy} describe how policies influence FDI in renewable energies. Accordingly, increasing FDI has supported the global expansion of renewable energies. The main source of FDI is Europe, in particular Germany and Spain, investing in renewable energy projects in the remaining European countries \cite{hanni2011foreign}. Such effects cannot be covered by the chosen model design. However, unlimited FDI could indeed be simulated by assuming the minimum WACC within the region of interest for all regions, meaning that all capital needed to make investment would be acquired in the region of minimum WACC. This resembles the setup of the homogeneous scenario and, thus, does not affect the general findings of this work. Furthermore, differences in WACC for renewable energy are mostly caused by tariff-related risks which apply for foreign direct investment as well \cite{temperton2016reducing}. As mentioned in Sec. \ref{sec:WACC}, WACC reflects varying market conditions between regions. Changing amounts of investment should not alter this. Concerning the generation and storage technologies available for expansion, we follow the conservative assumptions made in \citet{brown2018synergies}. In particular, other dispatchable zero emission technologies such as biomass and geothermal are not taken into account. Combined cycle gas turbines could in general be included. But since they would only partially replace OCGT, keeping the share of variable renewable generation sources unchanged, the inclusion would not affect the overall findings. Consequently, we excluded them in order to keep the model setup as simple as possible and the computation time appropriate as recommended by \citet{decarolis2017formalizing}.
%-------------------------------------------------------------------------------
\section{Conclusion}
\label{sec:conclusion}
%-------------------------------------------------------------------------------
Because power system modeling is an important tool for policy advise and system planning, the validity of the underlying assumptions is of crucial importance. Additionally considering the overall importance and urgency of establishing low-emission power systems to tackle climate change and the fact that no homogenization in WACC has been observed in Europe, we stress that the regional inhomogeneity in WACC needs to be taken into account in future studies or should at least be considered as a potential scenario. Furthermore, the uncertainty of the input parameters in power system models should be respected more carefully. Future work might focus on investigating the effects of this uncertainty on the outcome of power system optimization models.

%---------------------------------------------------------------
\section*{Acknowledgment}
%---------------------------------------------------------------
A. Kies is supported by the NetAllok project, which is funded by the German Federal Ministry for Economic Affairs and Energy (BMWi). The authors thank Jan Diettrich, Lüder von Bremen and Philipp Böttcher for helpful comments and fruitful discussions.

%---------------------------------------------------------------
\section*{Author contributions}
%---------------------------------------------------------------
B.~U.~Schyska prepared and executed the model runs. All authors contributed to designing the research, discussing the results and writing of the paper.

%---------------------------------------------------------------
\section*{Declaration of interests}
%---------------------------------------------------------------
The authors declare no conflict of interests.

%---------------------------------------------------------------
\section*{References}
%---------------------------------------------------------------
\bibliographystyle{model1-num-names}
\bibliography{sample.bib}

\begin{thebibliography}{61}
\expandafter\ifx\csname natexlab\endcsname\relax\def\natexlab#1{#1}\fi
\providecommand{\bibinfo}[2]{#2}
\ifx\xfnm\relax \def\xfnm[#1]{\unskip,\space#1}\fi
%Type = Misc
\bibitem[{{United Nations}(2015)}]{paris2015}
\bibinfo{author}{{United Nations}}, \bibinfo{title}{Paris agreement},
  \bibinfo{howpublished}{\url{https://unfccc.int/sites/default/files/english_paris_agreement.pdf}},
  \bibinfo{year}{2015}. \bibinfo{note}{Accessed: 2019-01-05}.
%Type = Article
\bibitem[{Brown et~al.(2018)Brown, Schlachtberger, Kies, Schramm, and
  Greiner}]{brown2018synergies}
\bibinfo{author}{T.~Brown}, \bibinfo{author}{D.~Schlachtberger},
  \bibinfo{author}{A.~Kies}, \bibinfo{author}{S.~Schramm},
  \bibinfo{author}{M.~Greiner},
\newblock \bibinfo{title}{Synergies of sector coupling and transmission
  reinforcement in a cost-optimised, highly renewable european energy system},
\newblock \bibinfo{journal}{Energy} \bibinfo{volume}{160}
  (\bibinfo{year}{2018}) \bibinfo{pages}{720--739}.
%Type = Article
\bibitem[{Usher and Strachan(2012)}]{usher2012critical}
\bibinfo{author}{W.~Usher}, \bibinfo{author}{N.~Strachan},
\newblock \bibinfo{title}{Critical mid-term uncertainties in long-term
  decarbonisation pathways},
\newblock \bibinfo{journal}{Energy Policy} \bibinfo{volume}{41}
  (\bibinfo{year}{2012}) \bibinfo{pages}{433--444}.
%Type = Article
\bibitem[{Ming et~al.(2017)Ming, Liu, Guo, Zhang, Feng, and
  Wang}]{ming2017optimizing}
\bibinfo{author}{B.~Ming}, \bibinfo{author}{P.~Liu}, \bibinfo{author}{S.~Guo},
  \bibinfo{author}{X.~Zhang}, \bibinfo{author}{M.~Feng},
  \bibinfo{author}{X.~Wang},
\newblock \bibinfo{title}{Optimizing utility-scale photovoltaic power
  generation for integration into a hydropower reservoir by incorporating
  long-and short-term operational decisions},
\newblock \bibinfo{journal}{Applied Energy} \bibinfo{volume}{204}
  (\bibinfo{year}{2017}) \bibinfo{pages}{432--445}.
%Type = Article
\bibitem[{Heide et~al.(2010)Heide, Von~Bremen, Greiner, Hoffmann, Speckmann,
  and Bofinger}]{heide2010seasonal}
\bibinfo{author}{D.~Heide}, \bibinfo{author}{L.~Von~Bremen},
  \bibinfo{author}{M.~Greiner}, \bibinfo{author}{C.~Hoffmann},
  \bibinfo{author}{M.~Speckmann}, \bibinfo{author}{S.~Bofinger},
\newblock \bibinfo{title}{Seasonal optimal mix of wind and solar power in a
  future, highly renewable europe},
\newblock \bibinfo{journal}{Renewable Energy} \bibinfo{volume}{35}
  (\bibinfo{year}{2010}) \bibinfo{pages}{2483--2489}.
%Type = Article
\bibitem[{Thomaidis et~al.(2016)Thomaidis, Santos-Alamillos, Pozo-V{\'a}zquez,
  and Usaola-Garc{\'\i}a}]{thomaidis2016optimal}
\bibinfo{author}{N.~S. Thomaidis}, \bibinfo{author}{F.~J. Santos-Alamillos},
  \bibinfo{author}{D.~Pozo-V{\'a}zquez},
  \bibinfo{author}{J.~Usaola-Garc{\'\i}a},
\newblock \bibinfo{title}{Optimal management of wind and solar energy
  resources},
\newblock \bibinfo{journal}{Computers \& Operations Research}
  \bibinfo{volume}{66} (\bibinfo{year}{2016}) \bibinfo{pages}{284--291}.
%Type = Article
\bibitem[{Francois et~al.(2016)Francois, Borga, Creutin, Hingray, Raynaud, and
  Sauterleute}]{francois2016complementarity}
\bibinfo{author}{B.~Francois}, \bibinfo{author}{M.~Borga},
  \bibinfo{author}{J.-D. Creutin}, \bibinfo{author}{B.~Hingray},
  \bibinfo{author}{D.~Raynaud}, \bibinfo{author}{J.-F. Sauterleute},
\newblock \bibinfo{title}{Complementarity between solar and hydro power:
  Sensitivity study to climate characteristics in northern-italy},
\newblock \bibinfo{journal}{Renewable energy} \bibinfo{volume}{86}
  (\bibinfo{year}{2016}) \bibinfo{pages}{543--553}.
%Type = Article
\bibitem[{Jurasz and Ciapa{\l}a(2017)}]{jurasz2017integrating}
\bibinfo{author}{J.~Jurasz}, \bibinfo{author}{B.~Ciapa{\l}a},
\newblock \bibinfo{title}{Integrating photovoltaics into energy systems by
  using a run-off-river power plant with pondage to smooth energy exchange with
  the power gird},
\newblock \bibinfo{journal}{Applied energy} \bibinfo{volume}{198}
  (\bibinfo{year}{2017}) \bibinfo{pages}{21--35}.
%Type = Article
\bibitem[{Jurasz(2017)}]{jurasz2017modeling}
\bibinfo{author}{J.~Jurasz},
\newblock \bibinfo{title}{Modeling and forecasting energy flow between national
  power grid and a solar--wind--pumped-hydroelectricity (pv--wt--psh) energy
  source},
\newblock \bibinfo{journal}{Energy conversion and management}
  \bibinfo{volume}{136} (\bibinfo{year}{2017}) \bibinfo{pages}{382--394}.
%Type = Article
\bibitem[{Jurasz et~al.(2018)Jurasz, Dabek, Ka{\'z}mierczak, Kies, and
  Wdowikowski}]{jurasz2018large}
\bibinfo{author}{J.~Jurasz}, \bibinfo{author}{P.~B. Dabek},
  \bibinfo{author}{B.~Ka{\'z}mierczak}, \bibinfo{author}{A.~Kies},
  \bibinfo{author}{M.~Wdowikowski},
\newblock \bibinfo{title}{Large scale complementary solar and wind energy
  sources coupled with pumped-storage hydroelectricity for lower silesia
  (poland)},
\newblock \bibinfo{journal}{Energy} \bibinfo{volume}{161}
  (\bibinfo{year}{2018}) \bibinfo{pages}{183--192}.
%Type = Article
\bibitem[{Santos-Alamillos et~al.(2015)Santos-Alamillos, Pozo-V{\'a}zquez,
  Ruiz-Arias, Von~Bremen, and Tovar-Pescador}]{santos2015combining}
\bibinfo{author}{F.~Santos-Alamillos}, \bibinfo{author}{D.~Pozo-V{\'a}zquez},
  \bibinfo{author}{J.~Ruiz-Arias}, \bibinfo{author}{L.~Von~Bremen},
  \bibinfo{author}{J.~Tovar-Pescador},
\newblock \bibinfo{title}{Combining wind farms with concentrating solar plants
  to provide stable renewable power},
\newblock \bibinfo{journal}{Renewable Energy} \bibinfo{volume}{76}
  (\bibinfo{year}{2015}) \bibinfo{pages}{539--550}.
%Type = Article
\bibitem[{Jurasz et~al.(2018)Jurasz, Beluco, and Canales}]{jurasz2018impact}
\bibinfo{author}{J.~Jurasz}, \bibinfo{author}{A.~Beluco},
  \bibinfo{author}{F.~A. Canales},
\newblock \bibinfo{title}{The impact of complementarity on power supply
  reliability of small scale hybrid energy systems},
\newblock \bibinfo{journal}{Energy} \bibinfo{volume}{161}
  (\bibinfo{year}{2018}) \bibinfo{pages}{737--743}.
%Type = Article
\bibitem[{Rodriguez et~al.(2014)Rodriguez, Becker, Andresen, Heide, and
  Greiner}]{rodriguez2014transmission}
\bibinfo{author}{R.~A. Rodriguez}, \bibinfo{author}{S.~Becker},
  \bibinfo{author}{G.~B. Andresen}, \bibinfo{author}{D.~Heide},
  \bibinfo{author}{M.~Greiner},
\newblock \bibinfo{title}{Transmission needs across a fully renewable european
  power system},
\newblock \bibinfo{journal}{Renewable Energy} \bibinfo{volume}{63}
  (\bibinfo{year}{2014}) \bibinfo{pages}{467--476}.
%Type = Article
\bibitem[{Kies et~al.(2016)Kies, Schyska, and von Bremen}]{kies2016curtailment}
\bibinfo{author}{A.~Kies}, \bibinfo{author}{B.~U. Schyska},
  \bibinfo{author}{L.~von Bremen},
\newblock \bibinfo{title}{Curtailment in a highly renewable power system and
  its effect on capacity factors},
\newblock \bibinfo{journal}{Energies} \bibinfo{volume}{9}
  (\bibinfo{year}{2016}) \bibinfo{pages}{510}.
%Type = Article
\bibitem[{Cao et~al.(2018)Cao, Metzdorf, and Birbalta}]{cao2018incorporating}
\bibinfo{author}{K.-K. Cao}, \bibinfo{author}{J.~Metzdorf},
  \bibinfo{author}{S.~Birbalta},
\newblock \bibinfo{title}{Incorporating power transmission bottlenecks into
  aggregated energy system models},
\newblock \bibinfo{journal}{Sustainability} \bibinfo{volume}{10}
  (\bibinfo{year}{2018}) \bibinfo{pages}{e1916}.
%Type = Article
\bibitem[{Steinke et~al.(2013)Steinke, Wolfrum, and Hoffmann}]{steinke2013grid}
\bibinfo{author}{F.~Steinke}, \bibinfo{author}{P.~Wolfrum},
  \bibinfo{author}{C.~Hoffmann},
\newblock \bibinfo{title}{Grid vs. storage in a 100\% renewable europe},
\newblock \bibinfo{journal}{Renewable Energy} \bibinfo{volume}{50}
  (\bibinfo{year}{2013}) \bibinfo{pages}{826--832}.
%Type = Article
\bibitem[{Weitemeyer et~al.(2015)Weitemeyer, Kleinhans, Vogt, and
  Agert}]{weitemeyer2015integration}
\bibinfo{author}{S.~Weitemeyer}, \bibinfo{author}{D.~Kleinhans},
  \bibinfo{author}{T.~Vogt}, \bibinfo{author}{C.~Agert},
\newblock \bibinfo{title}{Integration of renewable energy sources in future
  power systems: The role of storage},
\newblock \bibinfo{journal}{Renewable Energy} \bibinfo{volume}{75}
  (\bibinfo{year}{2015}) \bibinfo{pages}{14--20}.
%Type = Article
\bibitem[{Dunn et~al.(2011)Dunn, Kamath, and Tarascon}]{dunn2011electrical}
\bibinfo{author}{B.~Dunn}, \bibinfo{author}{H.~Kamath}, \bibinfo{author}{J.-M.
  Tarascon},
\newblock \bibinfo{title}{Electrical energy storage for the grid: a battery of
  choices},
\newblock \bibinfo{journal}{Science} \bibinfo{volume}{334}
  (\bibinfo{year}{2011}) \bibinfo{pages}{928--935}.
%Type = Article
\bibitem[{Weitemeyer et~al.(2016)Weitemeyer, Kleinhans, Wienholt, Vogt, and
  Agert}]{weitemeyer2016european}
\bibinfo{author}{S.~Weitemeyer}, \bibinfo{author}{D.~Kleinhans},
  \bibinfo{author}{L.~Wienholt}, \bibinfo{author}{T.~Vogt},
  \bibinfo{author}{C.~Agert},
\newblock \bibinfo{title}{A european perspective: potential of grid and storage
  for balancing renewable power systems},
\newblock \bibinfo{journal}{Energy Technology} \bibinfo{volume}{4}
  (\bibinfo{year}{2016}) \bibinfo{pages}{114--122}.
%Type = Article
\bibitem[{Heide et~al.(2011)Heide, Greiner, Von~Bremen, and
  Hoffmann}]{heide2011reduced}
\bibinfo{author}{D.~Heide}, \bibinfo{author}{M.~Greiner},
  \bibinfo{author}{L.~Von~Bremen}, \bibinfo{author}{C.~Hoffmann},
\newblock \bibinfo{title}{Reduced storage and balancing needs in a fully
  renewable european power system with excess wind and solar power generation},
\newblock \bibinfo{journal}{Renewable Energy} \bibinfo{volume}{36}
  (\bibinfo{year}{2011}) \bibinfo{pages}{2515--2523}.
%Type = Article
\bibitem[{Palensky and Dietrich(2011)}]{palensky2011demand}
\bibinfo{author}{P.~Palensky}, \bibinfo{author}{D.~Dietrich},
\newblock \bibinfo{title}{Demand side management: Demand response, intelligent
  energy systems, and smart loads},
\newblock \bibinfo{journal}{IEEE transactions on industrial informatics}
  \bibinfo{volume}{7} (\bibinfo{year}{2011}) \bibinfo{pages}{381--388}.
%Type = Article
\bibitem[{Zerrahn and Schill(2015)}]{zerrahn2015representation}
\bibinfo{author}{A.~Zerrahn}, \bibinfo{author}{W.-P. Schill},
\newblock \bibinfo{title}{On the representation of demand-side management in
  power system models},
\newblock \bibinfo{journal}{Energy} \bibinfo{volume}{84} (\bibinfo{year}{2015})
  \bibinfo{pages}{840--845}.
%Type = Article
\bibitem[{Kies et~al.(2016)Kies, Schyska, and von Bremen}]{kies2016demand}
\bibinfo{author}{A.~Kies}, \bibinfo{author}{B.~U. Schyska},
  \bibinfo{author}{L.~von Bremen},
\newblock \bibinfo{title}{The demand side management potential to balance a
  highly renewable european power system},
\newblock \bibinfo{journal}{Energies} \bibinfo{volume}{9}
  (\bibinfo{year}{2016}) \bibinfo{pages}{955}.
%Type = Article
\bibitem[{Hirth and M{\"u}ller(2016)}]{hirth2016system}
\bibinfo{author}{L.~Hirth}, \bibinfo{author}{S.~M{\"u}ller},
\newblock \bibinfo{title}{System-friendly wind power: How advanced wind turbine
  design can increase the economic value of electricity generated through wind
  power},
\newblock \bibinfo{journal}{Energy Economics} \bibinfo{volume}{56}
  (\bibinfo{year}{2016}) \bibinfo{pages}{51--63}.
%Type = Article
\bibitem[{Chattopadhyay et~al.(2017)Chattopadhyay, Kies, Lorenz, von Bremen,
  and Heinemann}]{chattopadhyay2017impact}
\bibinfo{author}{K.~Chattopadhyay}, \bibinfo{author}{A.~Kies},
  \bibinfo{author}{E.~Lorenz}, \bibinfo{author}{L.~von Bremen},
  \bibinfo{author}{D.~Heinemann},
\newblock \bibinfo{title}{The impact of different pv module configurations on
  storage and additional balancing needs for a fully renewable european power
  system},
\newblock \bibinfo{journal}{Renewable Energy} \bibinfo{volume}{113}
  (\bibinfo{year}{2017}) \bibinfo{pages}{176--189}.
%Type = Article
\bibitem[{Lund and Kempton(2008)}]{lund2008integration}
\bibinfo{author}{H.~Lund}, \bibinfo{author}{W.~Kempton},
\newblock \bibinfo{title}{Integration of renewable energy into the transport
  and electricity sectors through v2g},
\newblock \bibinfo{journal}{Energy policy} \bibinfo{volume}{36}
  (\bibinfo{year}{2008}) \bibinfo{pages}{3578--3587}.
%Type = Misc
\bibitem[{comission(2018)}]{europeancommission}
\bibinfo{author}{E.~comission}, \bibinfo{title}{{Clean Energy for all
  Europeans}},
  \bibinfo{howpublished}{\url{https://ec.europa.eu/energy/en/topics/energy-strategy-and-energy-union/clean-energy-all-europeans}},
  \bibinfo{year}{2018}. \bibinfo{note}{Accessed: 2018-09-30}.
%Type = Article
\bibitem[{Zappa et~al.(2019)Zappa, Junginger, and van~den Broek}]{zappa2019100}
\bibinfo{author}{W.~Zappa}, \bibinfo{author}{M.~Junginger},
  \bibinfo{author}{M.~van~den Broek},
\newblock \bibinfo{title}{Is a 100\% renewable european power system feasible
  by 2050?},
\newblock \bibinfo{journal}{Applied Energy} \bibinfo{volume}{233}
  (\bibinfo{year}{2019}) \bibinfo{pages}{1027--1050}.
%Type = Article
\bibitem[{Noothout et~al.(2016)Noothout, Jager, Tesni{\`e}re, van Rooijen,
  Karypidis, Br{\"u}ckmann et~al.}]{noothout2016diacore}
\bibinfo{author}{P.~Noothout}, \bibinfo{author}{D.~d. Jager},
  \bibinfo{author}{L.~Tesni{\`e}re}, \bibinfo{author}{S.~van Rooijen},
  \bibinfo{author}{N.~Karypidis}, \bibinfo{author}{R.~Br{\"u}ckmann}, et~al.,
\newblock \bibinfo{title}{Diacore. the impact of risks in renewable energy
  investments and the role of smart policies. final report},
\newblock \bibinfo{journal}{Ecofys. Utrecht}  (\bibinfo{year}{2016}).
%Type = Article
\bibitem[{Egli et~al.(2018)Egli, Steffen, and Schmidt}]{egli2018dynamic}
\bibinfo{author}{F.~Egli}, \bibinfo{author}{B.~Steffen}, \bibinfo{author}{T.~S.
  Schmidt},
\newblock \bibinfo{title}{A dynamic analysis of financing conditions for
  renewable energy technologies},
\newblock \bibinfo{journal}{Nature Energy}  (\bibinfo{year}{2018}).
%Type = Article
\bibitem[{Klessmann et~al.(2013)Klessmann, Rathmann, de~Jager, Gazzo, Resch,
  Busch, and Ragwitz}]{klessmann2013policy}
\bibinfo{author}{C.~Klessmann}, \bibinfo{author}{M.~Rathmann},
  \bibinfo{author}{D.~de~Jager}, \bibinfo{author}{A.~Gazzo},
  \bibinfo{author}{G.~Resch}, \bibinfo{author}{S.~Busch},
  \bibinfo{author}{M.~Ragwitz},
\newblock \bibinfo{title}{Policy options for reducing the costs of reaching the
  european renewables target},
\newblock \bibinfo{journal}{Renewable Energy} \bibinfo{volume}{57}
  (\bibinfo{year}{2013}) \bibinfo{pages}{390--403}.
%Type = Article
\bibitem[{Temperton(2016)}]{temperton2016reducing}
\bibinfo{author}{I.~Temperton},
\newblock \bibinfo{title}{Reducing the cost of financing renewables in europe},
\newblock \bibinfo{journal}{Agora Energiewende}  (\bibinfo{year}{2016}).
%Type = Article
\bibitem[{Kitzing et~al.(2012)Kitzing, Mitchell, and
  Morthorst}]{kitzing2012renewable}
\bibinfo{author}{L.~Kitzing}, \bibinfo{author}{C.~Mitchell},
  \bibinfo{author}{P.~E. Morthorst},
\newblock \bibinfo{title}{Renewable energy policies in europe: Converging or
  diverging?},
\newblock \bibinfo{journal}{Energy Policy} \bibinfo{volume}{51}
  (\bibinfo{year}{2012}) \bibinfo{pages}{192--201}.
%Type = Misc
\bibitem[{Brückmann(2018)}]{brueckmann}
\bibinfo{author}{R.~Brückmann}, \bibinfo{title}{{What is the development of
  WACC for wind power in the 28 EU Member States -and why?}},
  \bibinfo{howpublished}{\url{https://www.strommarkttreffen.org/2018-03_Brueckmann_Development_of_WACC_for_wind_in_EU28.pdf}},
  \bibinfo{year}{2018}. \bibinfo{note}{Accessed: 2018-11-30}.
%Type = Article
\bibitem[{Hirth and Steckel(2016)}]{hirth2016role}
\bibinfo{author}{L.~Hirth}, \bibinfo{author}{J.~C. Steckel},
\newblock \bibinfo{title}{The role of capital costs in decarbonizing the
  electricity sector},
\newblock \bibinfo{journal}{Environ. Res. Lett.} \bibinfo{volume}{11}
  (\bibinfo{year}{2016}) \bibinfo{pages}{114010}.
%Type = Article
\bibitem[{Schlachtberger et~al.(2017)Schlachtberger, Brown, Schramm, and
  Greiner}]{schlachtberger2017benefits}
\bibinfo{author}{D.~P. Schlachtberger}, \bibinfo{author}{T.~Brown},
  \bibinfo{author}{S.~Schramm}, \bibinfo{author}{M.~Greiner},
\newblock \bibinfo{title}{The benefits of cooperation in a highly renewable
  european electricity network},
\newblock \bibinfo{journal}{Energy} \bibinfo{volume}{134}
  (\bibinfo{year}{2017}) \bibinfo{pages}{469--481}.
%Type = Article
\bibitem[{Schlott et~al.(2018)Schlott, Kies, Brown, Schramm, and
  Greiner}]{schlott2018impact}
\bibinfo{author}{M.~Schlott}, \bibinfo{author}{A.~Kies},
  \bibinfo{author}{T.~Brown}, \bibinfo{author}{S.~Schramm},
  \bibinfo{author}{M.~Greiner},
\newblock \bibinfo{title}{The impact of climate change on a cost-optimal highly
  renewable european electricity network},
\newblock \bibinfo{journal}{Applied Energy} \bibinfo{volume}{230}
  (\bibinfo{year}{2018}) \bibinfo{pages}{1645--1659}.
%Type = Article
\bibitem[{DeCarolis et~al.(2017)DeCarolis, Daly, Dodds, Keppo, Li, McDowall,
  Pye, Strachan, Trutnevyte, Usher, Winning, Yeh, and
  Zeyringer}]{decarolis2017formalizing}
\bibinfo{author}{J.~DeCarolis}, \bibinfo{author}{H.~Daly},
  \bibinfo{author}{P.~Dodds}, \bibinfo{author}{I.~Keppo},
  \bibinfo{author}{F.~Li}, \bibinfo{author}{W.~McDowall},
  \bibinfo{author}{S.~Pye}, \bibinfo{author}{N.~Strachan},
  \bibinfo{author}{E.~Trutnevyte}, \bibinfo{author}{W.~Usher},
  \bibinfo{author}{M.~Winning}, \bibinfo{author}{S.~Yeh},
  \bibinfo{author}{M.~Zeyringer},
\newblock \bibinfo{title}{Formalizing best practice for energy system
  optimization modelling},
\newblock \bibinfo{journal}{Applied energy} \bibinfo{volume}{194}
  (\bibinfo{year}{2017}) \bibinfo{pages}{184--198}.
%Type = Article
\bibitem[{Mavromatidis et~al.(2018)Mavromatidis, Orehounig, and
  Carmeliet}]{mavromatidis2018uncertainty}
\bibinfo{author}{G.~Mavromatidis}, \bibinfo{author}{K.~Orehounig},
  \bibinfo{author}{J.~Carmeliet},
\newblock \bibinfo{title}{Uncertainty and global sensitivity analysis for the
  optimal design of distributed energy systems},
\newblock \bibinfo{journal}{Applied Energy} \bibinfo{volume}{214}
  (\bibinfo{year}{2018}) \bibinfo{pages}{219--238}.
%Type = Article
\bibitem[{Moret et~al.(2017)Moret, Giron{\`e}s, Bierlaire, and
  Mar{\'e}chal}]{moret2017characterization}
\bibinfo{author}{S.~Moret}, \bibinfo{author}{V.~C. Giron{\`e}s},
  \bibinfo{author}{M.~Bierlaire}, \bibinfo{author}{F.~Mar{\'e}chal},
\newblock \bibinfo{title}{Characterization of input uncertainties in strategic
  energy planning models},
\newblock \bibinfo{journal}{Applied energy} \bibinfo{volume}{202}
  (\bibinfo{year}{2017}) \bibinfo{pages}{597--617}.
%Type = Techreport
\bibitem[{Wealer et~al.(2019)Wealer, Bauer, Göke, von Hirschhausen, and
  Kemfert}]{DIW_nuclear}
\bibinfo{author}{B.~Wealer}, \bibinfo{author}{S.~Bauer},
  \bibinfo{author}{L.~Göke}, \bibinfo{author}{C.~von Hirschhausen},
  \bibinfo{author}{C.~Kemfert}, \bibinfo{title}{High-priced and dangerous:
  nuclear power is not an option for the climate-friendly energy mix},
  \bibinfo{type}{Technical Report} \bibinfo{number}{DIW Weekly Report 30/2019},
  DIW, \bibinfo{year}{2019}.
%Type = Misc
\bibitem[{Lazard(2019)}]{lazard}
\bibinfo{author}{Lazard}, \bibinfo{title}{Lazards levelized cost of energy
  analysis, version 11.0}, \bibinfo{year}{2019}.
%Type = Techreport
\bibitem[{Kies et~al.(2016)Kies, Chattopadhyay, von Bremen, Lorenz, and
  Heinemann}]{kies2016restore}
\bibinfo{author}{A.~Kies}, \bibinfo{author}{K.~Chattopadhyay},
  \bibinfo{author}{L.~von Bremen}, \bibinfo{author}{E.~Lorenz},
  \bibinfo{author}{D.~Heinemann}, \bibinfo{title}{Restore 2050: Simulation of
  renewable feed-in for power system studies}, \bibinfo{type}{Technical
  Report}, University of Oldenburg, \bibinfo{year}{2016}.
%Type = Article
\bibitem[{Weber et~al.(2019)Weber, Heinrichs, Gillessen, Schumann, H{\"o}rsch,
  Brown, and Witthaut}]{weber2019counter}
\bibinfo{author}{J.~Weber}, \bibinfo{author}{H.~U. Heinrichs},
  \bibinfo{author}{B.~Gillessen}, \bibinfo{author}{D.~Schumann},
  \bibinfo{author}{J.~H{\"o}rsch}, \bibinfo{author}{T.~Brown},
  \bibinfo{author}{D.~Witthaut},
\newblock \bibinfo{title}{Counter-intuitive behaviour of energy system models
  under co2 caps and prices},
\newblock \bibinfo{journal}{Energy} \bibinfo{volume}{170}
  (\bibinfo{year}{2019}) \bibinfo{pages}{22--30}.
%Type = Article
\bibitem[{Brown et~al.(2017)Brown, H{\"o}rsch, and
  Schlachtberger}]{brown2017pypsa}
\bibinfo{author}{T.~Brown}, \bibinfo{author}{J.~H{\"o}rsch},
  \bibinfo{author}{D.~Schlachtberger},
\newblock \bibinfo{title}{Pypsa: Python for power system analysis},
\newblock \bibinfo{journal}{arXiv preprint arXiv:1707.09913}
  (\bibinfo{year}{2017}).
%Type = Article
\bibitem[{Modigliani and Miller(1958)}]{modigliani1958cost}
\bibinfo{author}{F.~Modigliani}, \bibinfo{author}{M.~H. Miller},
\newblock \bibinfo{title}{The cost of capital, corporation finance and the
  theory of investment},
\newblock \bibinfo{journal}{The American economic review} \bibinfo{volume}{48}
  (\bibinfo{year}{1958}) \bibinfo{pages}{261--297}.
%Type = Article
\bibitem[{Modigliani and Miller(1963)}]{modigliani1963corporate}
\bibinfo{author}{F.~Modigliani}, \bibinfo{author}{M.~H. Miller},
\newblock \bibinfo{title}{Corporate income taxes and the cost of capital: a
  correction},
\newblock \bibinfo{journal}{The American economic review} \bibinfo{volume}{53}
  (\bibinfo{year}{1963}) \bibinfo{pages}{433--443}.
%Type = Techreport
\bibitem[{Schr{\"o}der et~al.(2013)Schr{\"o}der, Kunz, Meiss, Mendelevitch, and
  Von~Hirschhausen}]{schroder2013current}
\bibinfo{author}{A.~Schr{\"o}der}, \bibinfo{author}{F.~Kunz},
  \bibinfo{author}{J.~Meiss}, \bibinfo{author}{R.~Mendelevitch},
  \bibinfo{author}{C.~Von~Hirschhausen}, \bibinfo{title}{Current and
  prospective costs of electricity generation until 2050},
  \bibinfo{type}{Technical Report}, Data Documentation, DIW,
  \bibinfo{year}{2013}.
%Type = Article
\bibitem[{Rienecker et~al.(2011)Rienecker, Suarez, Gelaro, Todling, Bacmeister,
  Liu, Bosilovich, Schubert, Takacs, Kim et~al.}]{Merra}
\bibinfo{author}{M.~M. Rienecker}, \bibinfo{author}{M.~J. Suarez},
  \bibinfo{author}{R.~Gelaro}, \bibinfo{author}{R.~Todling},
  \bibinfo{author}{J.~Bacmeister}, \bibinfo{author}{E.~Liu},
  \bibinfo{author}{M.~G. Bosilovich}, \bibinfo{author}{S.~D. Schubert},
  \bibinfo{author}{L.~Takacs}, \bibinfo{author}{G.-K. Kim}, et~al.,
\newblock \bibinfo{title}{Merra: Nasa's modern-era retrospective analysis for
  research and applications},
\newblock \bibinfo{journal}{Journal of Climate} \bibinfo{volume}{24}
  (\bibinfo{year}{2011}) \bibinfo{pages}{3624--3648}.
%Type = Article
\bibitem[{Cano et~al.(1986)Cano, Monget, Albuisson, Guillard, Regas, and
  Wald}]{Cano_1986}
\bibinfo{author}{D.~Cano}, \bibinfo{author}{J.-M. Monget},
  \bibinfo{author}{M.~Albuisson}, \bibinfo{author}{H.~Guillard},
  \bibinfo{author}{N.~Regas}, \bibinfo{author}{L.~Wald},
\newblock \bibinfo{title}{A method for the determination of the global solar
  radiation from meteorological satellite data},
\newblock \bibinfo{journal}{Solar Energy} \bibinfo{volume}{37}
  (\bibinfo{year}{1986}) \bibinfo{pages}{31--39}.
%Type = Inproceedings
\bibitem[{Hammer et~al.(1998)Hammer, Heinemann, Westerhellweg, Ineichen,
  Olseth, Skartveit, Dumortier, Fontoynont, Wald, Beyer et~al.}]{Hammer_1998}
\bibinfo{author}{A.~Hammer}, \bibinfo{author}{D.~Heinemann},
  \bibinfo{author}{A.~Westerhellweg}, \bibinfo{author}{P.~Ineichen},
  \bibinfo{author}{J.~Olseth}, \bibinfo{author}{A.~Skartveit},
  \bibinfo{author}{D.~Dumortier}, \bibinfo{author}{M.~Fontoynont},
  \bibinfo{author}{L.~Wald}, \bibinfo{author}{H.~Beyer}, et~al.,
\newblock \bibinfo{title}{Derivation of daylight and solar irradiance data from
  satellite observations},
\newblock in: \bibinfo{booktitle}{Proceedings 9th Conference on Satellite
  Meteorology and Oceanography}, pp. \bibinfo{pages}{747--750}.
%Type = Article
\bibitem[{Klucher(1979)}]{Klucher_1979}
\bibinfo{author}{T.~Klucher},
\newblock \bibinfo{title}{Evaluation of models to predict insolation on tilted
  surfaces},
\newblock \bibinfo{journal}{Solar Energy} \bibinfo{volume}{23}
  (\bibinfo{year}{1979}) \bibinfo{pages}{111--114}.
%Type = Misc
\bibitem[{Kies et~al.(2017)Kies, von Bremen, and Heinemann}]{kies_zenodo}
\bibinfo{author}{A.~Kies}, \bibinfo{author}{L.~von Bremen},
  \bibinfo{author}{D.~Heinemann}, \bibinfo{title}{Hydro energy inflow for power
  system studies}, \bibinfo{year}{2017}.
%Type = Article
\bibitem[{Dee et~al.(2011)Dee, Uppala, Simmons, Berrisford, Poli, Kobayashi,
  Andrae, Balmaseda, Balsamo, Bauer et~al.}]{dee2011era}
\bibinfo{author}{D.~Dee}, \bibinfo{author}{S.~Uppala},
  \bibinfo{author}{A.~Simmons}, \bibinfo{author}{P.~Berrisford},
  \bibinfo{author}{P.~Poli}, \bibinfo{author}{S.~Kobayashi},
  \bibinfo{author}{U.~Andrae}, \bibinfo{author}{M.~Balmaseda},
  \bibinfo{author}{G.~Balsamo}, \bibinfo{author}{P.~Bauer}, et~al.,
\newblock \bibinfo{title}{The era-interim reanalysis: Configuration and
  performance of the data assimilation system},
\newblock \bibinfo{journal}{Quarterly Journal of the Royal Meteorological
  Society} \bibinfo{volume}{137} (\bibinfo{year}{2011})
  \bibinfo{pages}{553--597}.
%Type = Article
\bibitem[{Kies et~al.(2016{\natexlab{a}})Kies, Schyska, and von
  Bremen}]{kies2016effect}
\bibinfo{author}{A.~Kies}, \bibinfo{author}{B.~U. Schyska},
  \bibinfo{author}{L.~von Bremen},
\newblock \bibinfo{title}{The effect of hydro power on the optimal distribution
  of wind and solar generation facilities in a simplified highly renewable
  european power system},
\newblock \bibinfo{journal}{Energy Procedia} \bibinfo{volume}{97}
  (\bibinfo{year}{2016}{\natexlab{a}}) \bibinfo{pages}{149--155}.
%Type = Techreport
\bibitem[{Kies et~al.(2016{\natexlab{b}})Kies, Chattopadhyay, Lorenz, Bremen,
  and Heinemann}]{restore}
\bibinfo{author}{A.~Kies}, \bibinfo{author}{K.~Chattopadhyay},
  \bibinfo{author}{E.~Lorenz}, \bibinfo{author}{L.~v. Bremen},
  \bibinfo{author}{D.~Heinemann}, \bibinfo{title}{RESTORE 2050 Work Package
  Report D12: Simulation of renewable feed-in for power system studies},
  \bibinfo{type}{Technical Report}, University of Oldenburg, Insitute of
  Physics, ForWind, \bibinfo{year}{2016}{\natexlab{b}}.
%Type = Inproceedings
\bibitem[{Tranberg et~al.(2018)Tranberg, Sch{\"a}fer, Brown, H{\"o}rsch, and
  Greiner}]{tranberg2018flow}
\bibinfo{author}{B.~Tranberg}, \bibinfo{author}{M.~Sch{\"a}fer},
  \bibinfo{author}{T.~Brown}, \bibinfo{author}{J.~H{\"o}rsch},
  \bibinfo{author}{M.~Greiner},
\newblock \bibinfo{title}{Flow-based analysis of storage usage in a low-carbon
  european electricity scenario},
\newblock in: \bibinfo{booktitle}{2018 15th International Conference on the
  European Energy Market (EEM)}, \bibinfo{organization}{IEEE}, pp.
  \bibinfo{pages}{1--5}.
%Type = Article
\bibitem[{Pawel(2014)}]{pawel2014cost}
\bibinfo{author}{I.~Pawel},
\newblock \bibinfo{title}{The cost of storage--how to calculate the levelized
  cost of stored energy (lcoe) and applications to renewable energy
  generation},
\newblock \bibinfo{journal}{Energy Procedia} \bibinfo{volume}{46}
  (\bibinfo{year}{2014}) \bibinfo{pages}{68--77}.
%Type = Techreport
\bibitem[{Kovacevic et~al.(2018)Kovacevic, Assa, Bonini, Calderon, Hsu,
  Lengfelder, Mukhopadhyay, Nayyar, Rivera, and Tapia}]{HDI2018}
\bibinfo{author}{M.~Kovacevic}, \bibinfo{author}{J.~Assa},
  \bibinfo{author}{A.~Bonini}, \bibinfo{author}{C.~Calderon},
  \bibinfo{author}{Y.-C. Hsu}, \bibinfo{author}{C.~Lengfelder},
  \bibinfo{author}{T.~Mukhopadhyay}, \bibinfo{author}{S.~Nayyar},
  \bibinfo{author}{C.~Rivera}, \bibinfo{author}{H.~Tapia},
  \bibinfo{title}{Human Development Indices and Indicators -- 2018 Statistical
  Update}, \bibinfo{type}{Technical Report}, United Nations Development
  Programme, \bibinfo{year}{2018}.
%Type = Article
\bibitem[{Wall et~al.(2019)Wall, Grafakos, Gianoli, and
  Stavropoulos}]{wall2019policy}
\bibinfo{author}{R.~Wall}, \bibinfo{author}{S.~Grafakos},
  \bibinfo{author}{A.~Gianoli}, \bibinfo{author}{S.~Stavropoulos},
\newblock \bibinfo{title}{Which policy instruments attract foreign direct
  investments in renewable energy?},
\newblock \bibinfo{journal}{Climate policy} \bibinfo{volume}{19}
  (\bibinfo{year}{2019}) \bibinfo{pages}{59--72}.
%Type = Article
\bibitem[{Hanni et~al.(2011)Hanni, van Giffen, Kr{\"u}ger, and
  Mirza}]{hanni2011foreign}
\bibinfo{author}{M.~S. Hanni}, \bibinfo{author}{T.~van Giffen},
  \bibinfo{author}{R.~Kr{\"u}ger}, \bibinfo{author}{H.~Mirza},
\newblock \bibinfo{title}{Foreign direct investment in renewable energy:
  Trends, drivers and determinants},
\newblock \bibinfo{journal}{Transnational corporations} \bibinfo{volume}{20}
  (\bibinfo{year}{2011}) \bibinfo{pages}{29--65}.

\end{thebibliography}

\listoffigures

\clearpage

%--------------------------------------------------------------------------------------
\renewcommand{\thesection}{S}%\arabic{section}}
\renewcommand{\thesubsection}{\thesection.\arabic{subsection}}

\renewcommand{\thetable}{S\arabic{table}}
\setcounter{table}{0}

\renewcommand{\thefigure}{S\arabic{figure}}
\setcounter{figure}{0}

%--------------------------------------------------------------------------------------
\section{Supplementary Material}
\label{sec:supplementary}
%--------------------------------------------------------------------------------------
\begin{table*}[!ht]
    \centering
    \caption{Nomenclature}
    \begin{tabular}{lp{.95\textwidth}}
        \hline \\
         & \\
        $n$, $s$, $t$, $l$ & indices for node, generation/storage type, time and transmission link  \\
         & \\
        $c_{n,s}$ & investment costs for carrier $s$ at node $n$ [EUR/MW] \\
        $\text{CAP}_{\text{CO}_2}$ & global limit on $\text{CO}_2$ emissions [tons] \\
        $\text{CAP}_F$ & global limit of the sum of all single transmission line capacities [MWkm] \\
        $c_l$ & investment costs of transmission capacities at link $l$ [EUR/MWkm] \\
        $d_{n,t}$ & demand at node $n$ and time $t$ [MWh] \\
        $e_{n,s}$ & $\text{CO}_2$ emissions of generators of technology $s$ at node $n$ [tons/MWh] \\
        $\eta_{0,s}$ & standing losses of storage units of technology $s$ [a.u.] \\
        $\eta_{n,s}$ & efficiencies of generators of technology $s$ at node $n$ [a.u.] \\
        $\tau_{n.s}$ & energy-to-power ratio of storage units of technology $s$ at node $n$ [hours] \\
        $\lambda$ & dual variables \\
        $F_l$ & transmission capacities of link $l$ [MW] \\
        $f_{l,t}$ & flows over link $l$ at time $t$ [MWh] \\
        $G_{n,s}$ & capacity of generators or storage units of technology $s$ at node $n$ [MW] \\
        $g_{n,s,t}$ & dispatch of generators or storage units of technology $s$ at node $n$ and time $t$ [MWh] \\
        $g^-_{n,s,t}$ & maximal power uptake of generators or storage units of technology $s$ at node $n$ and time $t$ in units of $G_{n,s}$, zero for generators, negative for storage units \\
        $\bar{g}_{n,s,t}$ & maximum power output of generators or storage units of technology $s$ at node $n$ and time $t$ in units of $G_{n,s}$ \\
        $K_{n,l}$ & incidence matrix of the network \\
        $L_l$ & length of link $l$ [km] \\
        $o_{n,s}$ & marginal costs of generation of technology $s$ at node $n$ [EUR/MWh] \\
        $soc_{n,s,t}$ & state of charge of storage of technology $s$ at node $n$ and time $t$ \\
         & \\
        \hline
    \end{tabular}
    \label{tab:nomenclature}
\end{table*}

\begin{figure}[!ht]
    \centering
    \includegraphics[width=.48\textwidth]{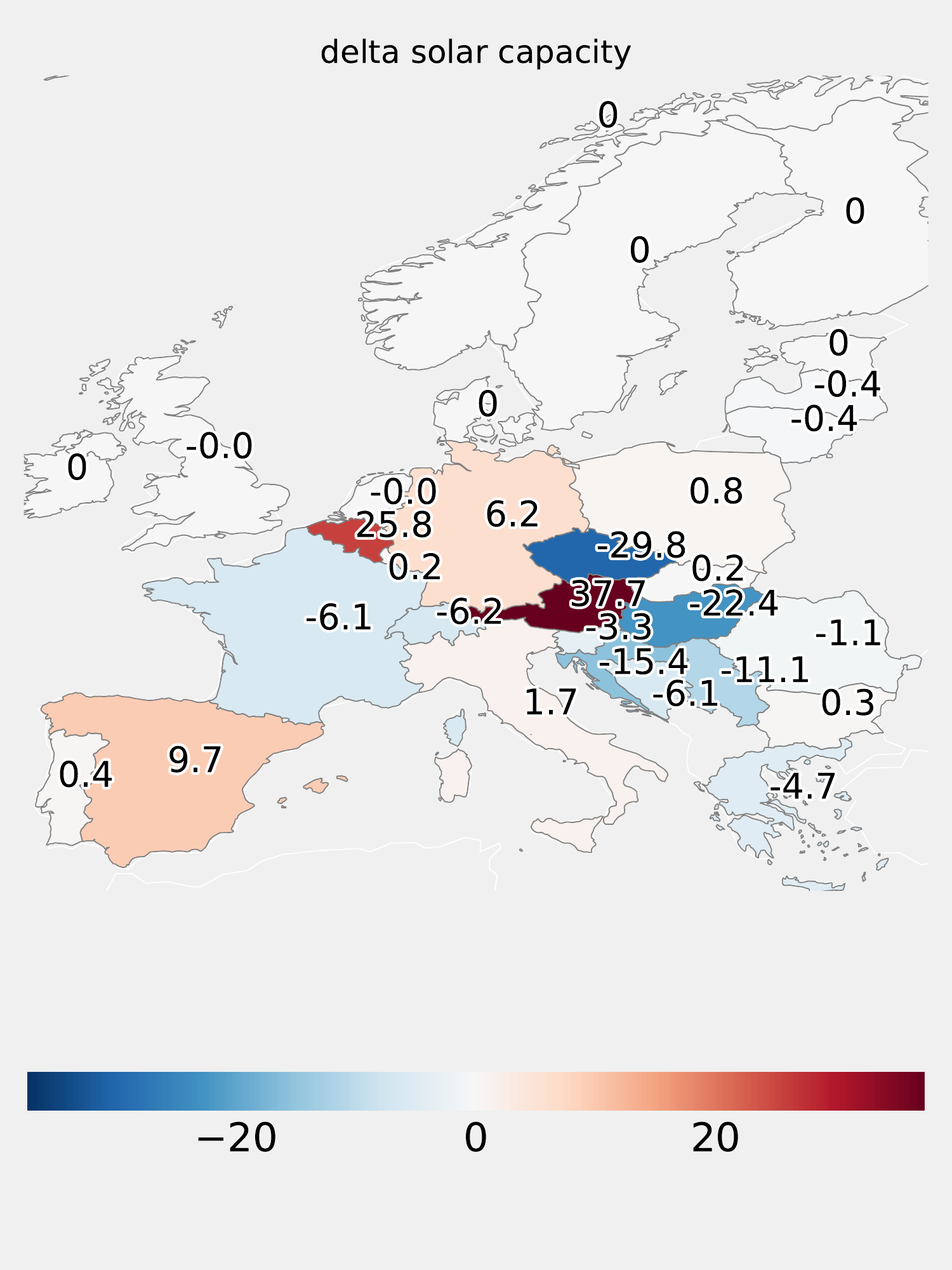}
    \includegraphics[width=.48\textwidth]{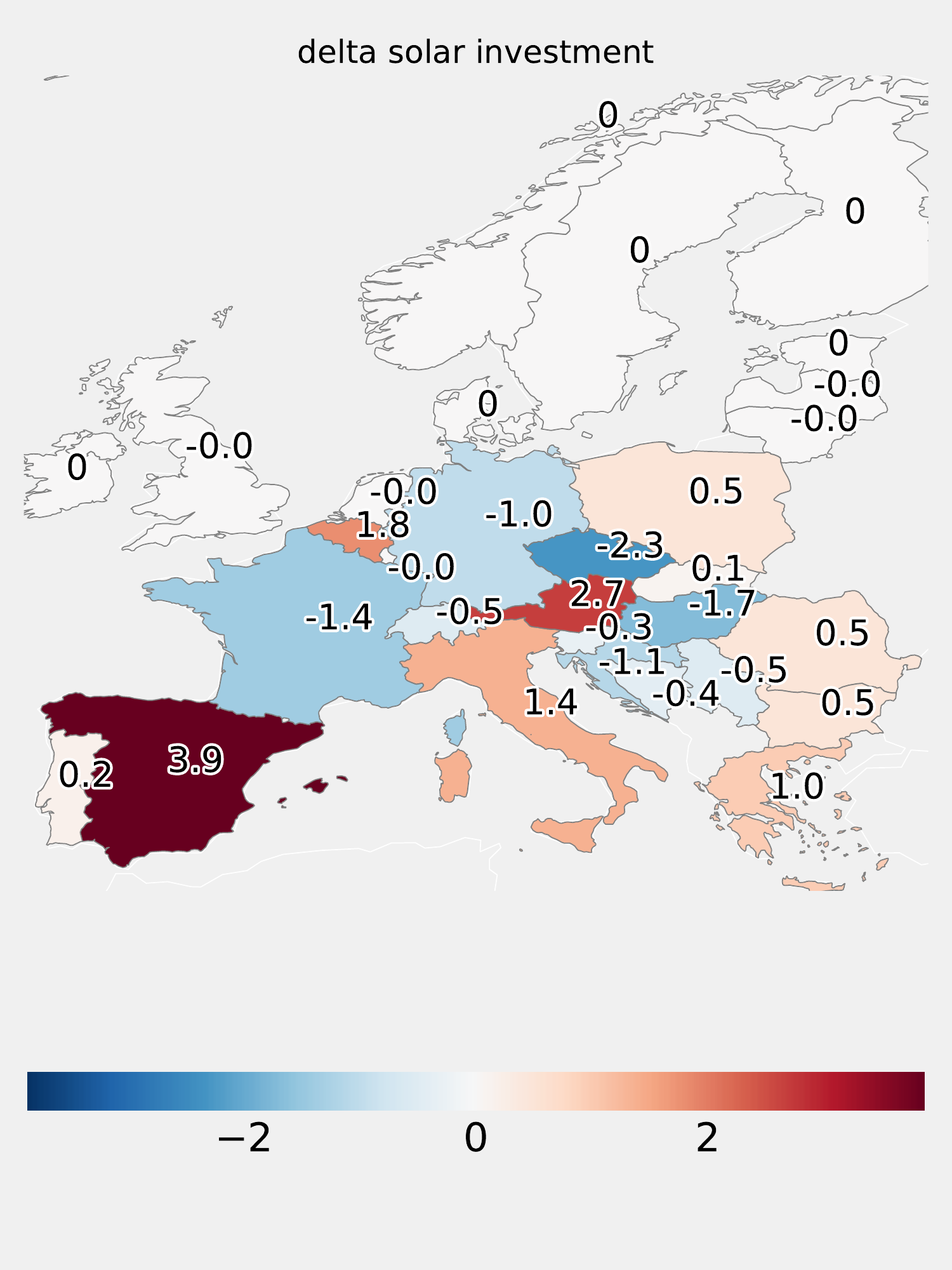}\\
    \caption{Regional differences in installed solar PV capacities [GW] (left) and investments [Bill. Euro] (right) between the homogeneous and the today scenario.}
    \label{fig:delta_solar}
\end{figure}
\begin{figure}[!ht]
    \centering
    \includegraphics[width=.48\textwidth]{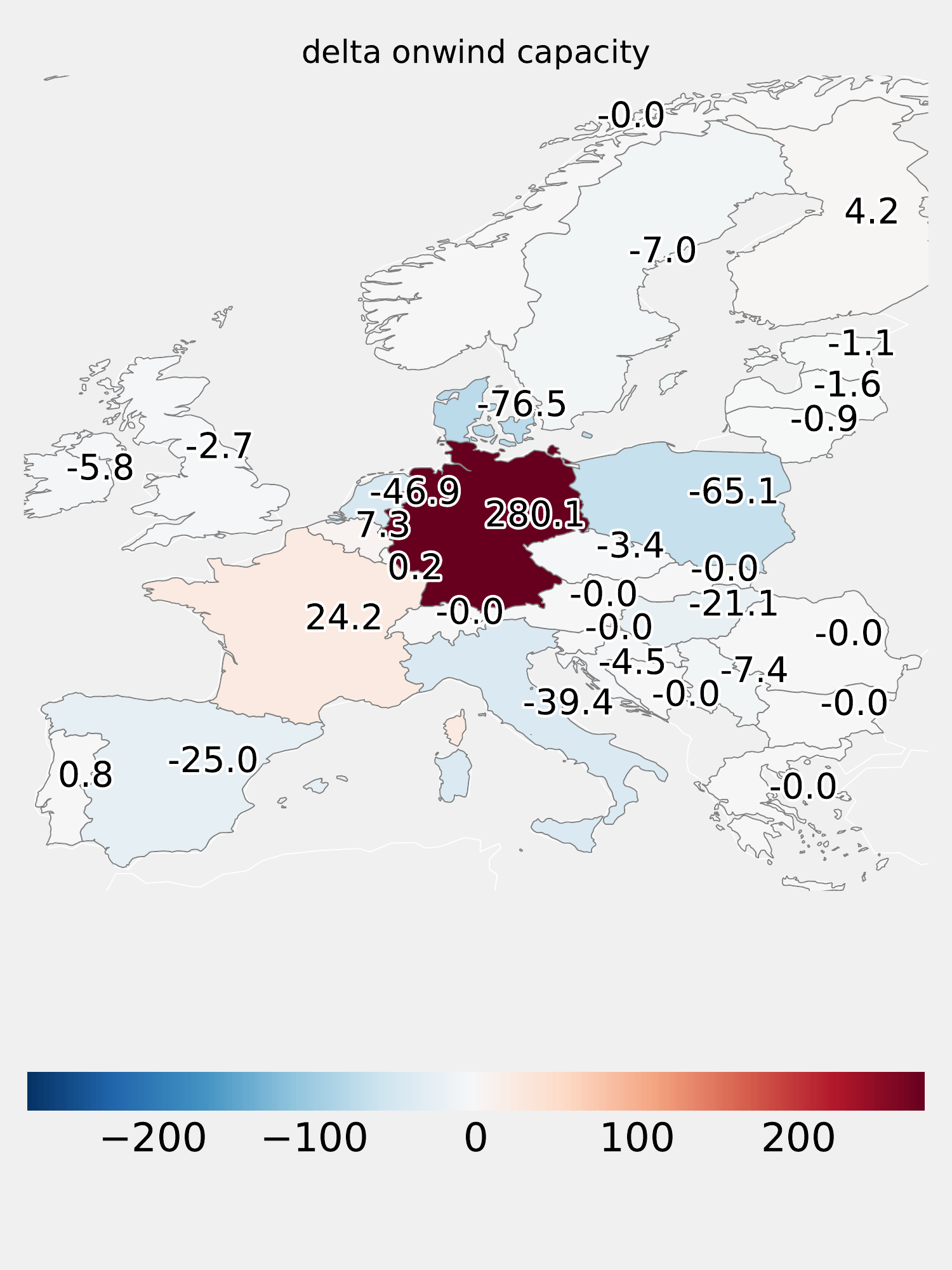}
    \includegraphics[width=.48\textwidth]{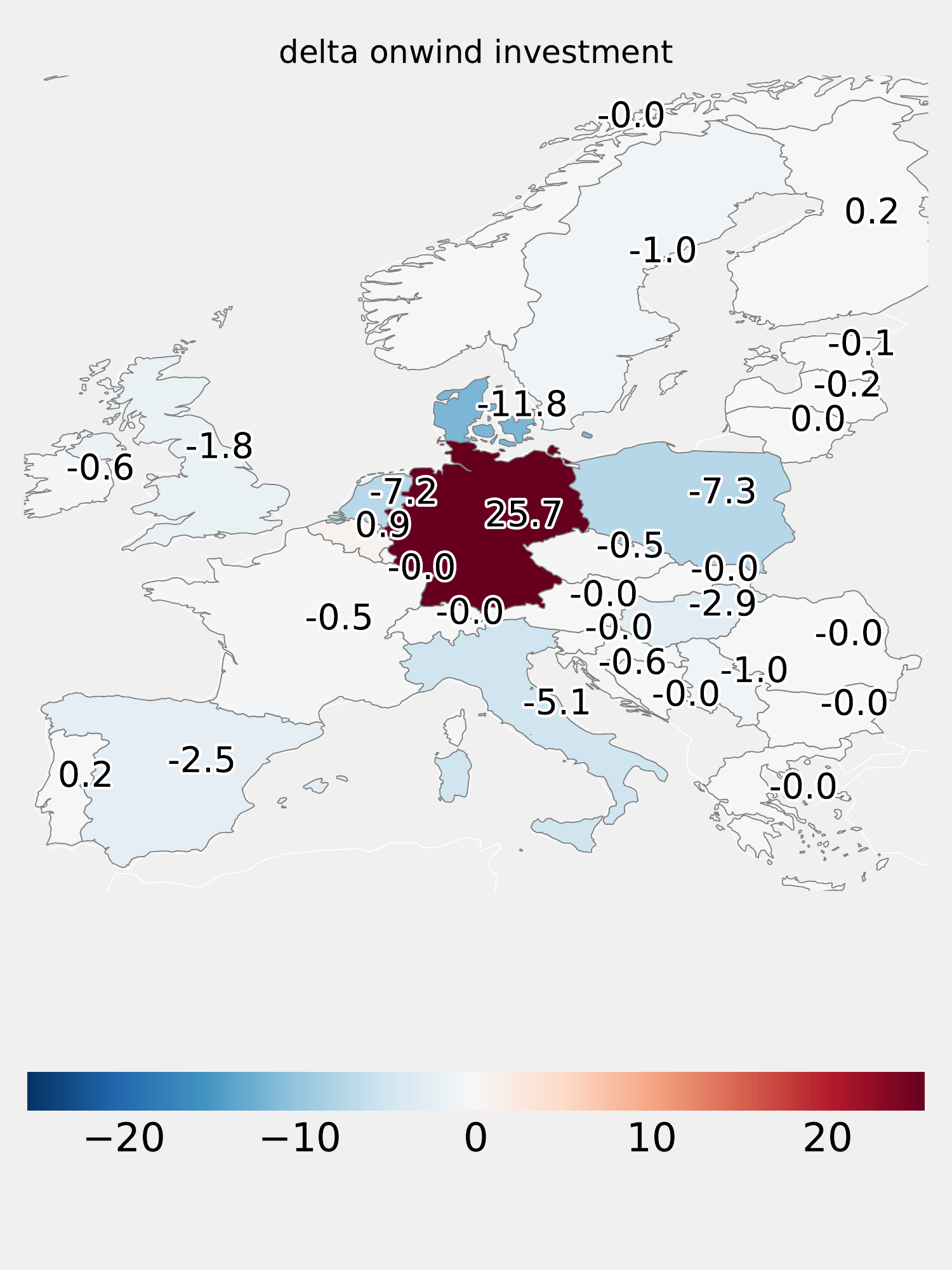}\\
    \caption{Regional differences in installed onshore wind capacities [GW] (left) and investments [Bill. Euro] (right) between the homogeneous and the today scenario.}
    \label{fig:delta_onwind}
\end{figure}
\begin{figure}[!ht]
    \centering
    \includegraphics[width=.48\textwidth]{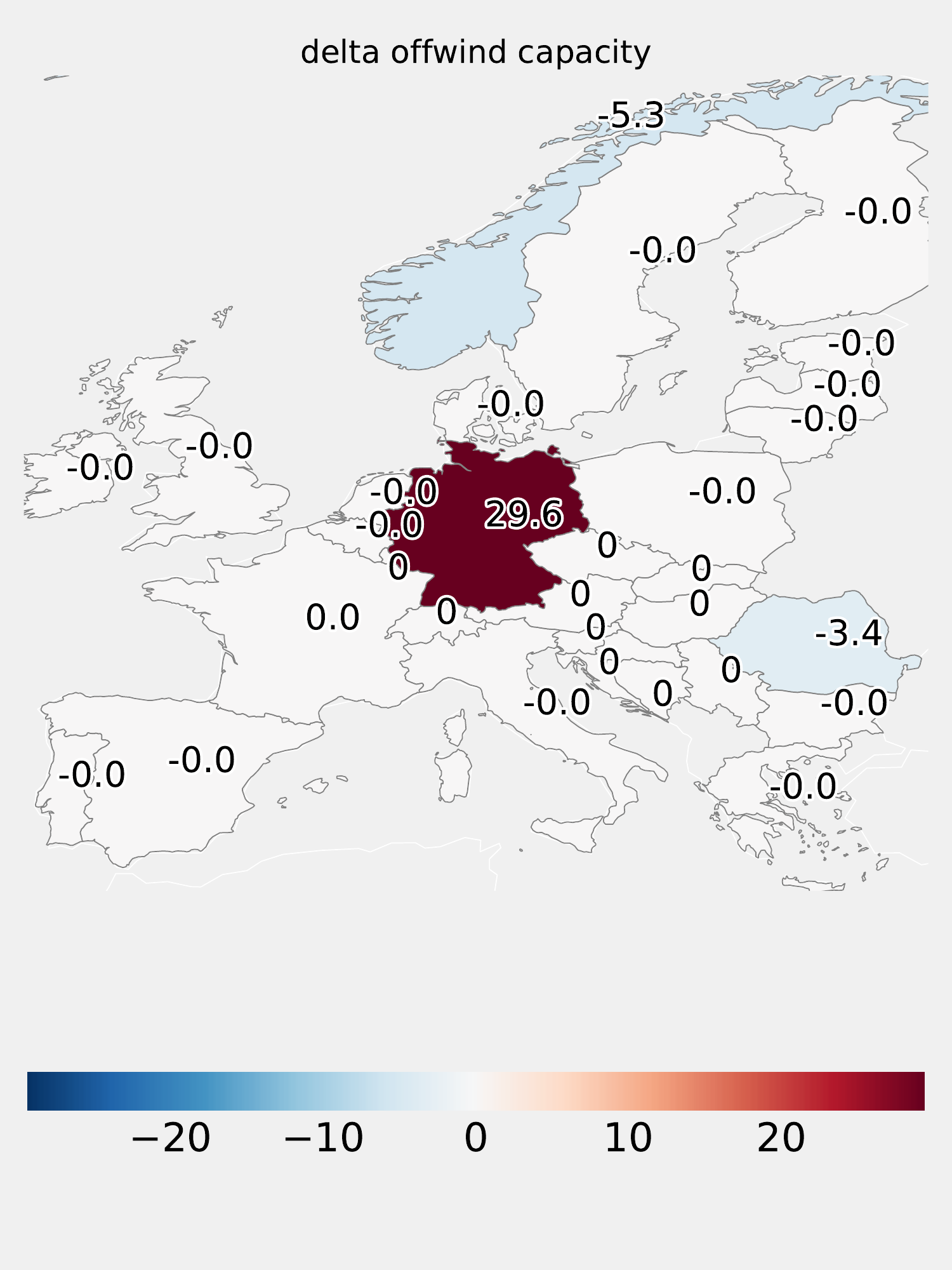}
    \includegraphics[width=.48\textwidth]{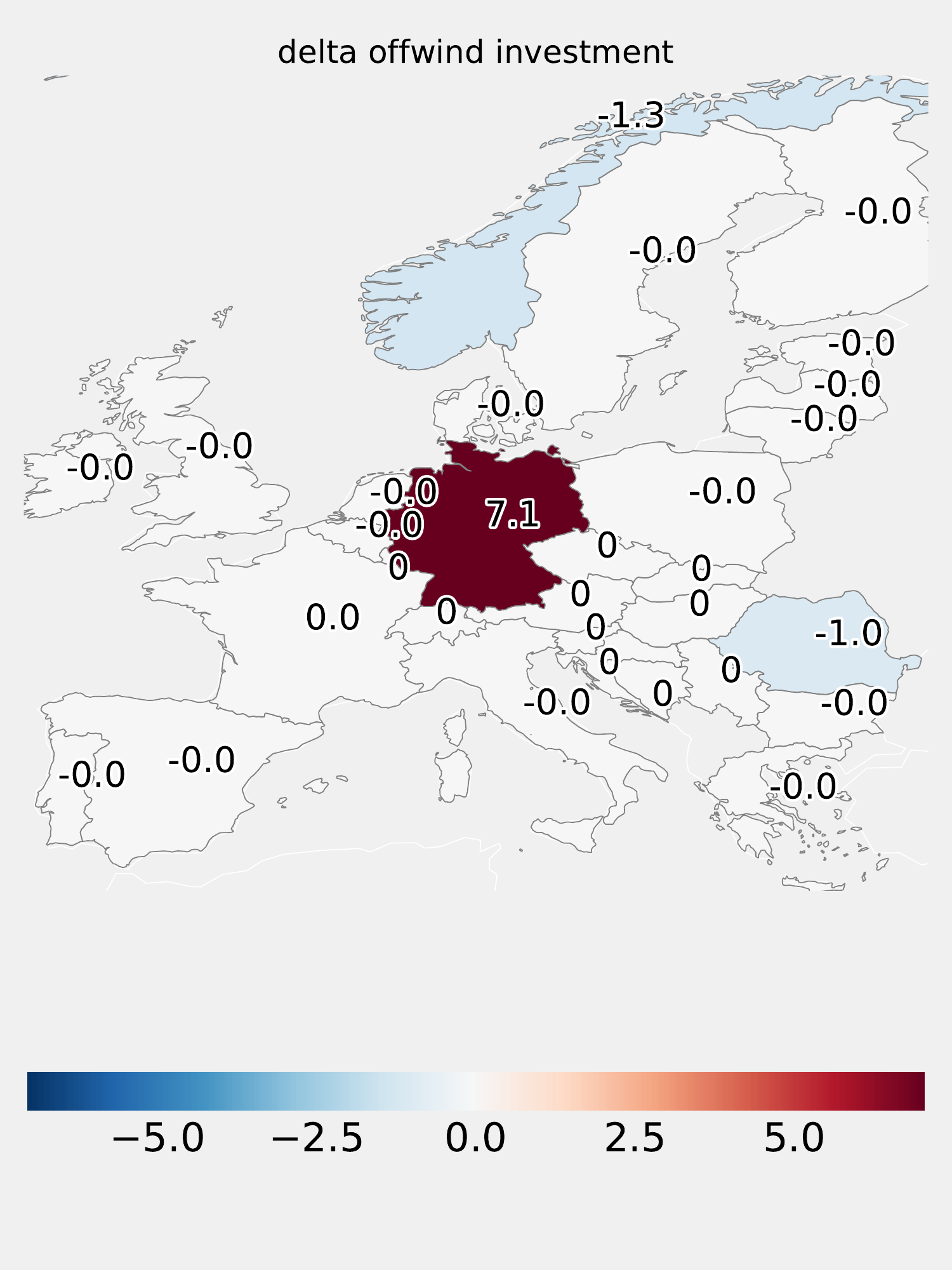}\\
    \caption{Regional differences in installed offshore wind capacities [GW] (left) and investments [Bill. Euro] (right) between the homogeneous and the today scenario.}
    \label{fig:delta_offwind}
\end{figure}
\begin{figure}
    \centering
    \includegraphics[width=.48\textwidth]{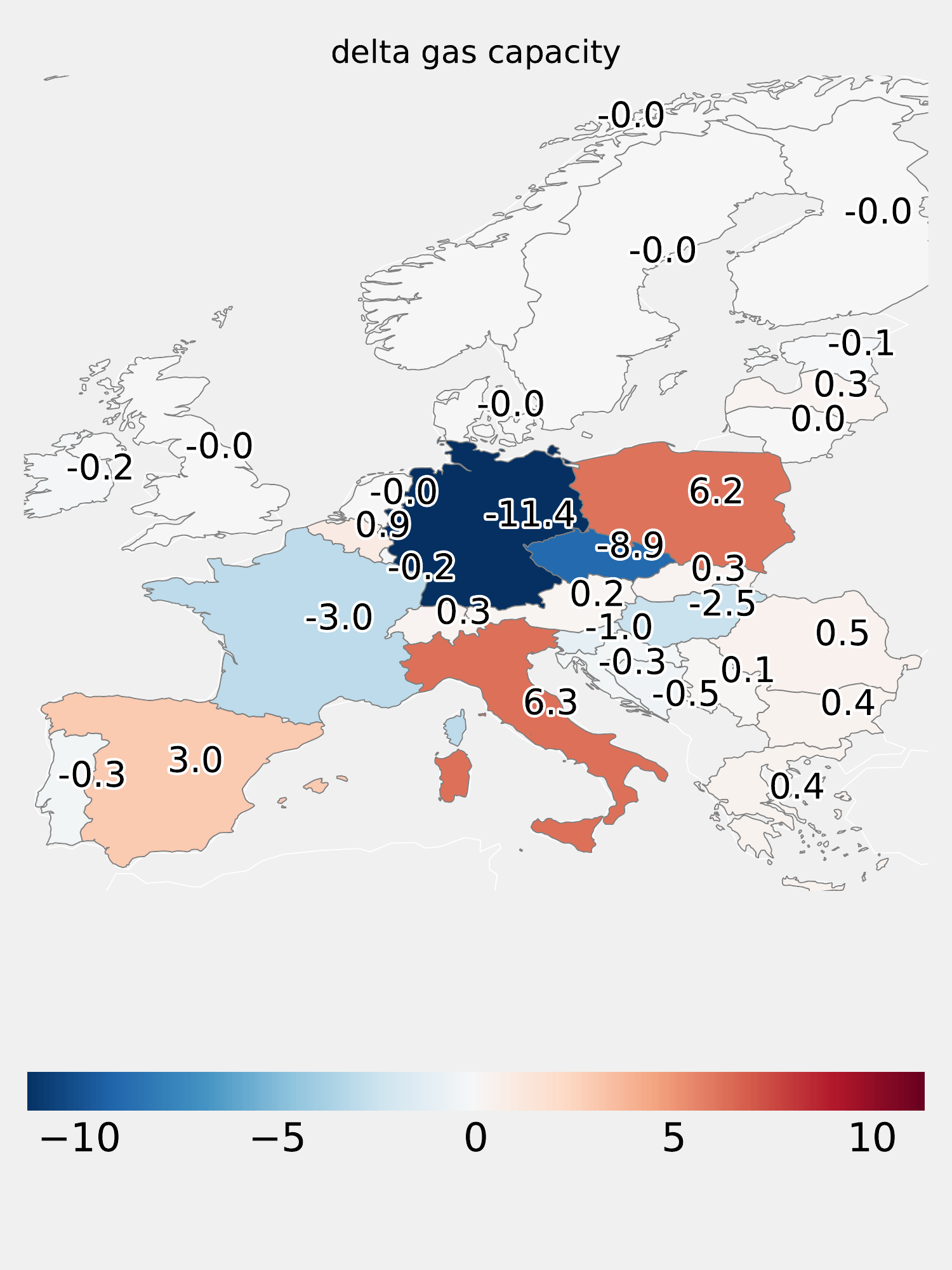}
    \includegraphics[width=.48\textwidth]{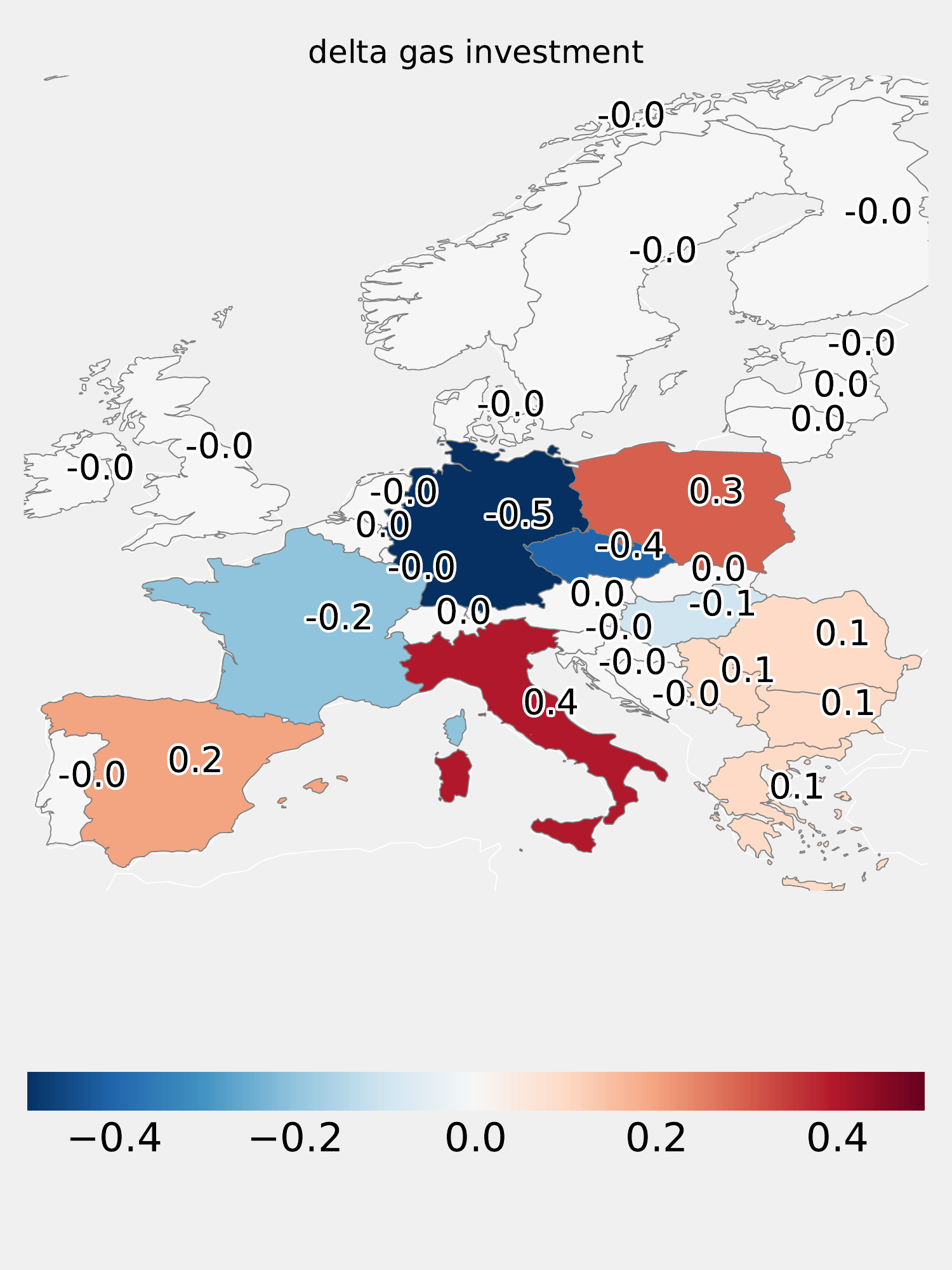}\\
    \caption{Regional differences in installed OCGT capacities [GW] (left) and investments [Bill. Euro] (right) between the homogeneous and the today scenario..}
    \label{fig:delta_gas}
\end{figure}

\end{document}